\documentclass[
        12pt,
        notitlepage, 
        tightenlines,
        eqsecnum,
        floats,
        floatfix,
        nofootinbib,
        amsmath,amssymb,
        aps,prd,
        superscriptaddress,
        longbibliography]{revtex4-2}

\usepackage[utf8]{inputenc}
\DeclareUnicodeCharacter{0302}{}
\usepackage{microtype}
\usepackage{comment}
\usepackage{mathtools}
\usepackage{graphicx}
\usepackage{caption}
\usepackage{subcaption}
\usepackage{float}
\usepackage{amssymb}
\usepackage{xcolor}
\usepackage[hidelinks]{hyperref}
\usepackage{cleveref}

\def\dd{{\rm d}}

\definecolor{newgreen}{rgb}{0.0, 0.75, 0.0}

\begin{document}

\title{Stellar collapse with pressure in effective loop quantum gravity}


\author{Luca Cafaro}
\email{lcafaro@fuw.edu.pl}
\affiliation{ Faculty of Physics, University of Warsaw, Pasteura 5, 02-093 Warsaw, Poland}

\author{Lorenzo Cipriani}
\email{lorenzo.cipriani@graduate.univaq.it}
\affiliation{Dipartimento di Scienze Fisiche e Chimiche, Università dell’Aquila, via Vetoio, I-67100, L’Aquila, Italy}
\affiliation{INFN, Laboratori Nazionali del Gran Sasso, I-67100 Assergi (AQ), Italy}

\author{Francesco Fazzini}
 \email{francesco.fazzini@unb.ca}
\affiliation{ Department of Mathematics and Statistics, University of New Brunswick, Fredericton, NB, Canada E3B 5A3}

\author{Farshid Soltani}
 \email{f.soltani@uw.edu.pl}
\affiliation{ Faculty of Physics, University of Warsaw, Pasteura 5, 02-093 Warsaw, Poland}


\begin{abstract}
\vspace*{.3cm}

We explore semiclassical stellar collapse scenarios with pressure within the framework of effective loop quantum gravity. The objective of this work is to generalize existent models of semiclassical dust collapse and examine the role of pressure in the formation of shell-crossing singularities in a semiclassical context. Numerical investigations show that the singularity characterizing the end state of any classical stellar collapse is here resolved by quantum gravitational effects and replaced by a bounce of the star. However, they also show that shell-crossing singularities remain a general feature of these models and that the inclusion of pressure does not alter the qualitative picture emerging from semiclassical models of inhomogeneous dust collapse. Given the absence of a black hole singularity and the possibility of extending spacetime in the future of the trapped region formed by gravitational collapse, the investigation of the causal structure of the spacetime describing the semiclassical collapse of a star is inevitably tied to a better understanding of the physics of these shell-crossing singularities.
\end{abstract}

\maketitle

\section{Introduction}

In the classical theory, the gravitational collapse of a spherically symmetric star generally results in the formation of a black hole. The collapse continues in the interior of the black hole until the entire star is collapsed, or focused, into a singular event of zero physical radius with infinite density. This signals a complete breakdown and loss of predictivity of the classical theory in the interior of a black hole. It is a common belief that quantum gravitational effects would alter this picture and prevent the formation of the singularity.

Loop quantum gravity (LQG) is mature enough to address this problem in a semiclassical setting, where leading quantum corrections to Einstein field equations are considered to obtain an effective model of spacetime. This semiclassical approach originated from loop quantum cosmology (LQC)~\cite{Ashtekar_2011}, where the symmetries of spacetime are imposed at the classical level and then this classical symmetry-reduced theory is quantized using the LQG algebra of holonomies of the connection and areas. It is then possible to show that the semiclassical limit of this quantum theory describes an effective spacetime model whose dynamics is given by an effective Hamiltonian. Furthermore, at least in the cosmological setting, the resulting effective equations provide a surprisingly accurate approximation to leading quantum corrections even in deep quantum regions~\cite{Rovelli_2014,Bojowald_2008,Diener_2014}. The same techniques have been then applied to spherically symmetric black hole spacetimes~\cite{ashtekar2023regularblackholesloop}, starting from the vacuum solution. It is only in recent years (with some notable exceptions~\cite{Bojowald:2008ja,Bojowald:2009ih}) that effective equations of motion have been used to study stellar collapse.

Without a complete fundamental theory, effective equations are not unique, and different choices can be made at the effective level. A choice of effective equations that is widely used in loop quantum cosmology and effective black hole spacetimes goes under the name of $\Bar{\mu}$-K-scheme (see e.g.~\cite{Ashtekar:2006wn,Singh_2013,Vandersloot:2006ws,Ashtekar:2006es}). We will also adopt this choice. It is however interesting to point out that covariant effective schemes were recently found in~\cite{Alonso-Bardaji:2021yls,zhang2024bhc,belfaqih2024,Alonso-Bardaji:2023qgu}. The relationship between all these schemes is still unclear.

Given its relative simplicity, the Oppenheimer-Snyder (OS) model has been the first spherically symmetric stellar collapse model that was studied using these techniques~\cite{BenAchour:2020mgu,BenAchour:2020gon,Kelly:2020lec,Lewandowski:2022zce,Han:2023wxg,Fazzini:2023scu,Bobula:2023kbo,Giesel_2023,Boldorini:2024bgg}. As expected, the classical dynamics ending in a singularity is replaced by a non-singular dynamics in which the star collapses, reaches a minimum radius where its energy density becomes Planckian, and then it ``bounces'' and starts to expand. There is still some disagreement about what happens at the boundary of the star at the moment of the bounce, as the dust field (not its physical energy density, which in OS is discontinuous from the start) develops a discontinuity there~\cite{Fazzini:2023ova}. However, as the curvature and the metric remain completely regular in that point~\cite{Fazzini:2023scu}, we believe nothing peculiar, besides the bounce, happens there.

A more realistic stellar collapse scenario is given by an inhomogeneous dust collapse, also known as the Lemaître-Tolman-Bondi (LTB) model. In the classical theory, a new general feature of stellar collapse is associated with this model. It introduces a new type of singularity where the local energy density diverges because two spherical matter shells try to cross each other due to the collapse dynamics. To distinguish these singularities from the ones discussed before, we will call the former shell-crossing singularities (SCS) and the latter black hole singularities. Spacetime singularities are generally divided into strong and weak singularities depending on their properties~\cite{Tipler1977,Clarke1985,Krolak1986}. A strong singularity is a singular region of spacetime where curvature diverges and any infalling object is necessarily crushed by tidal forces. A weak singularity is a singular region of spacetime where curvature may or may not diverge, but infalling objects are able to continue their trajectory unharmed beyond the singularity. While black hole singularities are strong singularities, shell-crossing singularities are weak. Namely, spacetime can be continued past a shell-crossing singularity and so the latter poses no harm to the predictivity of the theory~\cite{Newman_1986}.

This notwithstanding, shell-crossing singularities still imply a divergence in the energy density of the star and the curvature of spacetime. It is a widespread belief that either the inclusion of pressure (with a proper equation of state), the inclusion of quantum gravitational corrections, or the conjunction of the two, would resolve these singularities as well. One possibility is that shell-crossing singularities would turn into non-singular acoustic waves~\cite{Plebanski_Krasinski_2006}. If this is the case, it would be reasonable to assume that the somewhat singular dynamics in the future of the shell-crossing singularities, which can be found by looking for weak solutions to the equations of motion as done in~\cite{Nolan_2003}, gives a qualitatively accurate description of the non-singular dynamics of these acoustic waves.

In the classical theory, stellar collapse scenarios with pressure have been considered in~\cite{hagendas1974}. There it was shown that even with the inclusion of pressure, in the form of a perfect fluid with pressure being a function of only energy density, shell-crossing singularities remain a general feature of stellar collapse. This result seems to suggest that pressure alone is not enough to resolve shell-crossing singularities. An argument could however be made that the usual equation of state for perfect fluids where pressure is only a function of the energy density might not be the most reasonable one in an inhomogeneous setting~\cite[pp.~296,324]{Plebanski_Krasinski_2006}.

The LTB dust collapse scenario has been recently studied also in the context of semiclassical LQG effective models~\cite{Bojowald:2008ja,Bojowald:2009ih,Husain:2021ojz,Husain:2022gwp,giesel2023embeddinggeneralizedltbmodels,Giesel:2023hys,Fazzini:2023ova,Cipriani:2024nhx,Bobula:2024ywp,Bobula:2024chr,giesel2024rbh}. These analyses confirmed the resolution of the black hole singularity by quantum gravitational effects already seen in the Oppenheimer-Snyder model, but they also found formation of shell-crossing singularities as in the classical theory. In~\cite{Fazzini:2023ova} it was indeed shown that each initial continuous energy density profile of compact support develops shell-crossing singularities at most within a Planckian time from the quantum bounce of the core of the star. This shows that the quantum gravitational corrections included in these LQG effective models do not resolve shell-crossing singularities by themselves. In fact, the effective LQG models built so far seem in general to resolve strong singularities and ignore weak singularities~\cite{Singh_2009,Singh_2011,Giesel:2024mps}.

In this work, we investigate stellar collapse scenarios with pressure in the context of semiclassical LQG effective models. The effective equations of motion for spherically symmetric and inhomogeneous spacetimes with a perfect fluid in the generalized Painlevé-Gullstrand gauge were recently found in~\cite{Wilson-Ewing:2024uad}, where static solutions were also studied. We reformulate the model in the LTB gauge, which is more convenient for studying collapsing models, and generalize the effective equations of motion for the case of an anisotropic fluid with vanishing heat conduction and viscosity (\cref{sec:effectiveeqs}). We then explored the case of the collapse of an anisotropic fluid with vanishing radial pressure (\cref{TANG_PRES}). While certainly not realistic, this toy model allows us to investigate analytically the interplay of pressure and quantum corrections in a semiclassical setting. Armed with the insight developed with this toy model, we study the collapse of a perfect fluid with a linear equation of state (\cref{sec:perfect}). Given the complexity of the system of evolution equations for this scenario, this case must be investigated numerically. As expected, in both scenarios the black hole singularity that develops in the classical theory is resolved by quantum gravitational effects. However, shell-crossing singularities are still a general feature of these models, even for high pressures. This means that the conjunction of pressure and the quantum gravitational corrections included in LQG effective models is still not enough to resolve these (weak) singularities. In \cref{sec:vacuum} we briefly discuss the vacuum solutions of the effective model and we notice the existence of a bigger class of static solutions than previously thought. Finally, in \cref{sec:conclusions}, we comment on the results we obtained.

\section{Effective equations}
\label{sec:effectiveeqs}

\subsection{Generalized Painlevé-Gullstrand gauge}

The LQG effective equations of motion for spherically symmetric and inhomogeneous spacetimes with a perfect fluid have been recently derived in generalized Painlevé-Gullstrand coordinates~\cite{Wilson-Ewing:2024uad}. This section briefly outlines this construction and generalizes the results to anisotropic fluids.

Consider a spherically symmetric spacetime with a fluid having a 4-velocity field $u_\mu$ and zero rotation ($\omega_{\mu\nu}=0$). In terms of generalized Painlevé-Gullstrand coordinates, its metric takes the form
\begin{equation} 
\dd s^2=-N^2 \dd t^2 +\frac{1}{1+\varepsilon} (\dd r + N^r \dd t)^2 +r^2 \dd \Omega^2 , 
\label{PG1}
\end{equation}
with $N(t,r)$ and $N^r (t,r)$ being respectively the lapse function and the radial component of the shift vector (the other components being zero by symmetry), and $\varepsilon(t,r)>-1$. The defining properties of these coordinates are: (i) the radial coordinate $r$ is the areal coordinate satisfying $A = 4\pi r^2$, with $A$ being the proper area of the surfaces of constant $t$ and $r$; (ii) the hypersurfaces of constant $t$ are the hypersurfaces orthogonal to the irrotational flow of the fluid, namely $u_\mu \propto \partial_\mu t $. When the motion of the fluid is geodesic, it is possible to find a $t$ such that $u_\mu = - \partial_\mu t $, which means that the lapse function can be taken to be $N=1$. This happens for example when the fluid is dust, where the choice $N=1$ is usually called the dust-time gauge because, in the Hamiltonian framework, it is a result of the gauge fixing condition $\phi=t$, where $\phi$ is the dust field. But also in vacuum, in Schwarzschild spacetime, where $u^\mu$ is now associated with geodesic observers in radial motion. This is the case in which Painlevé-Gullstrand coordinates were originally formulated, with the additional requirement of $\varepsilon=0$ (which in Schwarzschild spacetime corresponds to the choice of radial geodesic observers that start at rest at infinity). When the motion of the fluid is not geodesic, as in our case, $u_\mu = - N \partial_\mu t $ and the lapse function cannot be put to $N=1$.

In terms of the Ashtekar-Barbero variables that the LQG effective models are based on, the metric of a spherically symmetric spacetime in areal gauge reads~\cite{Kelly:2020uwj}
\begin{equation} \label{PG}
\dd s^2=-N^2 \dd t^2 +\left(\frac{E^b}{r}\right)^2 (\dd r + N^r \dd t)^2 +r^2 \dd \Omega^2 , 
\end{equation}
where $E^{b}$ is the angular component of the symmetry-reduced densitized triad and the radial component of the shift vector is given by
\begin{equation}
 N^r=-\frac{Nr}{\gamma \sqrt{\Delta}}\sin \left(\frac{\sqrt{\Delta }b}{r} \right) \cos \left( \frac{\sqrt{\Delta} b}{r}\right). 
 \label{shiftvector}
\end{equation}
In the last expression, $b$ is the angular component of the symmetry-reduced extrinsic curvature and the variable canonically conjugate to $E^{b}$, $\gamma$ is the Barbero-Immirzi parameter, and $\Delta \sim \ell^2_{\mathrm{Pl}}$ is the area gap, the smallest non-zero eigenvalue of the LQG area operator. The only remaining constraint is the effective Hamiltonian constraint $\mathcal H \approx 0$, with~\cite{Wilson-Ewing:2024uad}
\begin{equation}
\mathcal{H}=\left[-\frac{E^b}{2G \gamma^2 \Delta r}\partial_r \left( r^3 \sin^2\frac{\sqrt{\Delta}b}{r} \right)-\frac{1}{2G}\left(\frac{E^b}{r}-\frac{r}{E^b} \right)  +\frac{r}{G}\partial_r \left( \frac{r}{E^b}\right)+ \mathcal{H}_m  \right] ,   
\end{equation}
where $\mathcal{H}_m$ is the Hamiltonian density of the fluid. The effective Hamiltonian  $\mathcal H$ is the generator of the semiclassical dynamics that include quantum gravitational corrections, in the form of holonomy corrections, motivated by loop quantum gravity.

Before exploring the dynamical evolution generated by the effective scalar constraint we need to specify the class of fluids we want to study. A perfect fluid is a fluid with no heat conduction, no viscosity, and that is isotropic in its rest frame. We keep the first two properties but allow anisotropy in the form of different radial and angular, or tangential, pressures in the rest frame of the fluid. This generalization serves only a technical purpose in this article, as it will allow us to study the interplay of pressure and quantum corrections in the simplified albeit unphysical case of vanishing radial pressure, where a few analytical results are attainable. However, anisotropies of this kind might actually play an important role in the structure of some stars~\cite{cadogan2024}. The stress-energy tensor $T_{\mu\nu}$ of these anisotropic fluids can be covariantly written as
\begin{equation}
T^{\mu}_{\nu}=\rho \, u^{\mu}u_{\nu}+\Pi \,e^{\mu}e_{\nu} +\Sigma \left(\delta^{\mu}_{\nu}+u^{\mu}u_{\nu}-e^{\mu}e_{\nu}\right),
\label{Tmunu}
\end{equation}
where $u_\mu = -N \partial_\mu t$ is the 4-velocity of the fluid, $e^{\mu} = \sqrt{1+\varepsilon}\, \delta^\mu_r$ is the unit spacelike vector defining the direction of anisotropy of the fluid, and $\rho (t,r)$, $\Pi(t,r)$, and $\Sigma(t,r)$ are respectively the energy density, radial pressure, and tangential pressure in the rest frame of the fluid. The energy density $\rho$ and the radial pressure $\Pi$ are related to the Hamiltonian $H_m=\int \dd r N \mathcal{H}_m$ of the fluid in the following way~\cite{Singh_2009} (see~\cite{Darryl1986} for an expression of $H_m$ in an anisotropic context)
\begin{align}
 &\rho=\frac{\mathcal{H}_m}{V}=\frac{\mathcal{H}_m}{4 \pi r E^b}~, \label{rho1}\\
&\Pi=-\frac{\delta H_m}{\delta V}=-\frac{1}{4 \pi r}\frac{\delta H_m}{\delta E^b} \label{pressure} \, ,
\end{align}
where $V$ is the spatial volume element. Notice that the previous relations are formally equal to the classical ones. This is because the fluid part of the effective Hamiltonian constraint is not polymerized.

Making the change of variable
\begin{equation}
E^b=\frac{r}{\sqrt{1+\varepsilon}} \label{varepsilon}
\end{equation}
to go back to the variables used in \cref{PG1}, we can now study the effective LQG dynamics of our semiclassical model in the same variables normally used in the classical theory. First of all, given the expression for the energy density of the fluid given in \cref{rho1}, the imposition of the effective Hamiltonian constraint $\mathcal{H}\approx 0$ gives the following on-shell expression for $\rho$:
\begin{equation} \label{rho_def}
\rho=\frac{1}{8 \pi G r^2} \partial_r \left(\frac{1}{\gamma^2 \Delta}r^3 \sin^2 \frac{\sqrt{\Delta}b}{r}-\varepsilon r   \right) .
\end{equation}
In the classical theory, the Einstein field equations tell us that the Misner-Sharp mass
\begin{equation} \label{M_def}
 m(r,t)=4\pi \int_0 ^r \rho ~\Tilde{r}^2 \dd \Tilde{r}=\frac{1}{2G}\left(\frac{1}{\gamma^2 \Delta} r^3 \sin^2 \frac{\sqrt{\Delta} b}{r}-\varepsilon r \right),
\end{equation}
gives the total quantity of mass-energy inside radius $r$. We will see that the same remains true also at the effective level. The effective equations of motion (EOMs) for our dynamical variables read
\begin{align}
    \frac{\partial b}{\partial t} &=\left\{b, H \right\} =-\frac{N}{2 \gamma \Delta r}\partial_r \left(r^3 \sin^2 \frac{\sqrt{\Delta}b}{r} \right)+\frac{\gamma N \varepsilon}{2r}+\gamma \left( 1 +\varepsilon \right) (\partial_r N)-4 \pi r \gamma G \Pi  N ,\label{b_dot} \\
    \frac{\partial \varepsilon}{\partial t} &=\left\{\varepsilon,H\right\}=-\frac{r}{\gamma \sqrt{\Delta}}\sin\frac{\sqrt{\Delta}b}{r}\cos\frac{\sqrt{\Delta}b}{r}\left[N \partial_r \varepsilon - 2(1+\varepsilon)(\partial_r N)\right] \label{eps_dot}  . 
\end{align}
Another useful effective equation is the evolution equation for the gravitational mass, which, from \cref{M_def}, is given by
\begin{equation}
   \frac{\partial m}{\partial t} =-\frac{N r}{\gamma \sqrt{\Delta}}\sin\frac{\sqrt{\Delta}b}{r}\cos\frac{\sqrt{\Delta}b}{r}(\partial_r m+4 \pi r^2 \Pi) \label{M_dot}  .
\end{equation}
Furthermore, from \cref{M_def} we have
\begin{equation}
    \sin^2{\frac{\sqrt{\Delta}b}{r}}=\frac{8 \pi \gamma^2 G \Delta}{r^3} \int_0^r \dd \tilde{r} \, \tilde{r}^2 \rho + \gamma^2 \Delta \frac{\varepsilon}{r^2}  = \frac{\gamma^2 \Delta}{r^2} \left( \frac{2 G m}{r}+\varepsilon \right),
\end{equation}
which combined with \cref{shiftvector} allows us to express $N^r$ as
\begin{equation} 
     \big(N^r\big)^2= N^2 \bigg( \frac{2 G m}{r} +\varepsilon\bigg)
     \,\bigg[ 1-\frac{\gamma^2 \Delta}{r^2} \left( \frac{2 G m}{r}+\varepsilon \right) \bigg] \, ,
     \label{veryimportant}
\end{equation}
where the sign of $N^r$ discerns between a locally contracting ($N^r>0$) or expanding ($N^r<0$) solution. \Cref{veryimportant} provides an equivalent way to impose the effective Hamiltonian constraint $\mathcal{H}\approx0$ that will prove to be very useful.

These equations can be used to study the dynamics of all the dynamical metric variables except the lapse function $N$. An equation for the latter is given by the energy-momentum conservation law $\nabla_\mu T^{\mu\nu}=0$, which still holds at the effective level because the polymerization does not affect the matter part of Einstein field equations. The projection of this equation on the plane orthogonal to $u^\mu$ gives
\begin{equation}
    \frac{\partial_r N}{N} =\frac{1}{\rho+ \Pi }\left[\frac{2}{r}(\Sigma-\Pi)- \partial_r \Pi  \right] . 
    \label{lapse1}
\end{equation}
This equation, together with \cref{b_dot,eps_dot} and the equations of state (EOS) for the pressures $\Pi$ and $\Sigma$, give a complete set of equations for all the unknowns of the problem. Solving these equations of motion in the areal gauge is however quite challenging. As already noticed in~\cite{Fazzini:2023ova} for the dust case, the Lemaître-Tolman-Bondi gauge provides a set of equations that is simpler to solve numerically. For this reason, we will now reformulate the problem in this gauge.

\subsection{Lemaître-Tolman-Bondi gauge}

A disclaimer is now in order. The effective LQG models constructed using the $\Bar{\mu}$-K-scheme~\cite{Ashtekar:2006wn,Singh_2013,Vandersloot:2006ws,Ashtekar:2006es} are not ``covariant'', in the sense that solutions of the effective equations corresponding to different classical gauges are not necessarily related by a change of coordinates at the semiclassical level. This is a known limitation of the $\Bar{\mu}$-K-scheme~\cite{Bojowald_2021}. To make contact with previous results in the effective collapse of a star, we choose to start with the effective theory developed in the generalized Painlevé-Gullstrand gauge in the previous section and then to make a change of coordinates to go in the Lemaître-Tolman-Bondi gauge. This model is not necessarily the same model one would obtain by developing the effective theory directly in the Lemaître-Tolman-Bondi gauge. However, the two models were in fact shown to be equivalent in the case of dust~\cite{Giesel:2023hys}. Given that the matter contribution to the constraints is not modified by quantum correction, we expect this equivalence to hold also in the more general case discussed here. A rigorous proof of this property is left for future work.

The defining properties of the Lemaître-Tolman-Bondi coordinates are: (I) the spatial coordinates are comoving with the fluid, which means $u^\mu \propto \delta^\mu_0$; (II) the hypersurfaces of constant $t$ are the hypersurfaces orthogonal to the irrotational flow of the fluid, namely $u_\mu \propto \partial_\mu t $ as in the Painlevé-Gullstrand coordinates. Condition (II) tells us that the time coordinate $t$ is the same in both gauges. Given the change of coordinate $r=r(t,R)$ into a new radial coordinate $R$, condition (I) implies $\partial_t r= - N^r$. The line element in LTB coordinates $(t, R, \theta, \varphi)$ then reads
\begin{equation} \label{LTB}
\dd s^2=-N^2(t,R) \dd t^2 +\frac{\left( \partial_R r \right)^2}{1+\varepsilon(t,R)} \dd R^2 +r^2(t,R) \dd \Omega^2 . 
\end{equation}
Notice that condition (I) does not impose any constraint on $\partial_R r$. This translates in a reparametrization invariance of the metric under change of comoving coordinate $R$. We can use this freedom to assume $r(t_0,R)=R$ in the following.


Now we can rewrite the effective equations of motion of the previous section in the new coordinates. Using the shorthand notion $\dot{X}(t,R)=\partial_t X(t,R)$ and $X^{\prime}(t,R)=\partial_R X(t,R)$, we have
\begin{align}
& \dot{m}(t,R)=-\frac{4 \pi N r^3 \Pi }{\gamma \sqrt{\Delta}}\sin\frac{\sqrt{\Delta}b}{r}\cos\frac{\sqrt{\Delta}b}{r}=-4\pi \Pi r^2 \Dot{r} , \label{massevolution}
\\
& \dot{\varepsilon}(t,R)=2(1+\varepsilon)\frac{N^{\prime} r }{\gamma \sqrt{\Delta} r^{\prime}}
\sin\frac{\sqrt{\Delta}b}{r}\cos\frac{\sqrt{\Delta}b}{r}=2 (1+\varepsilon)\frac{N^{\prime}}{N}\frac{\dot{r}}{r^{\prime}} \, ,
\end{align}
where we used $\dot{r}=-N^r$ in the second equality of both expressions. The evolution equation for the mass $m$ in \cref{massevolution} gives an energy-conservation equation: the time derivative of the mass inside $R$ equals the volume-work done on the system. This tells us that, as in the classical theory, the Misner-Sharp mass $m(t,R)$ gives the total mass-energy inside the shell $R$ at time $t$. In the new coordinates, the mass function reads
\begin{equation} \label{mass_def}
m(t,R)=4\pi \int_0^{R} \rho(t,\tilde{R}) \, r(t,\tilde{R})^2 \,r'(t, \tilde{R}) \dd \tilde{R} ,
\end{equation}
which means that the energy density is given by
\begin{equation}
   \rho=\frac{m'}{ 4 \pi r^2 r'} \, .
   \label{rho}
\end{equation}
This is the same expression one finds in the classical theory. Therefore, as in the classical case, the system admits in principle black hole singularities where $r(R,t)=0$ and $m'(t,R) \neq 0$, shell-crossing singularities where $r'(t,R)=0$ and $m'(t,R) \neq 0$, and (``regular'') shell-crossings where $r'=0$ but $m'=0$ as well and $\rho$ remains finite. We will see how the effective dynamics avoids the formation of black hole singularities but still leads to shell-crossing singularities for generic initial data.

The energy-momentum conservation law $\nabla_{\mu} T^{\mu\nu}=0 $ gives equations for lapse and energy density:
\begin{align}
\frac{N'}{N} &=\frac{1}{\rho+ \Pi }\left[\frac{2r'}{r}(\Sigma-\Pi)-\Pi ' \right] , \label{lapse}\\
\Dot{\rho} &=-\left(\frac{\Dot{r}'}{r'}-\frac{1}{2} \frac{\dot{\varepsilon}}{1+\varepsilon} \right) \left(\rho + \Pi\right) - 2 \frac{\dot{r}}{r} \left(\rho +\Sigma\right) .
\label{rhodot}
\end{align}
A dynamical equation for the areal radius $r(t,R)$ is given by the effective Hamiltonian constraint in the form of \cref{veryimportant}. Finally, we need to complete the set of dynamical equations by providing two equations of state $f(\Pi, \rho)=0$ and $g(\Sigma, \rho)=0$ for the pressures.

The set of differential equations we need to solve to specify the dynamics of semiclassical stellar collapse in LTB coordinates is then:
\begin{align} 
\Dot{r}^2= &\, N^2 \bigg( \frac{2 G m}{r} +\varepsilon\bigg)
     \,\bigg[ 1-\frac{\gamma^2 \Delta}{r^2} \left( \frac{2 G m}{r}+\varepsilon \right) \bigg],
\label{r} \\
\Dot{m}= & -4\pi \Pi\, r^2 \Dot{r} , 
\label{mass}\\
\Dot{\varepsilon}=&\, \frac{(1+\varepsilon)}{r'}\frac{2\Dot{r}}{\rho+ \Pi }\left[\frac{2r'}{r}(\Sigma-\Pi)-\Pi ' \right], 
\label{epsilon}\\
\Dot{\rho} =& -\left(\frac{\Dot{r}'}{r'}-\frac{1}{2} \frac{\dot{\varepsilon}}{1+\varepsilon} \right) \left(\rho + \Pi\right) - 2 \frac{\dot{r}}{r} \left(\rho +\Sigma\right) ,
\label{rhodotset}\\
\frac{N'}{N} =&\, \frac{1}{\rho+ \Pi }\left[\frac{2r'}{r}(\Sigma-\Pi)-\Pi ' \right] ,
\label{Nset}\\
 \Dot{\Pi}=&\: \frac{\delta \Pi}{\delta \rho} \dot{\rho} , 
\label{Piset}\\
 \Dot{\Sigma}=&\: \frac{\delta \Sigma}{\delta \rho}  \dot{\rho} .
\label{sigma}
\end{align} 
The mass $m$ and the energy density $\rho$ are not independent variables, so we do not need dynamical equations for both of them. However, as they both give valuable insight into the physics of the problem, we will keep both of them remembering that they are not independent. So we have a system of 6 independent coupled PDEs in 6 unknowns, $r(t,R)$, $\rho(t,R)$, $\varepsilon(t,R)$, $N(t,R)$, $\Pi(t,R)$, and $\Sigma(t,R)$. Once the initial data is given, this system of equations can be solved numerically and the physics of the semiclassical collapse of a star with pressure can be completely specified. Interestingly, except for \cref{r} where quantum gravitational $\Delta$-corrections are explicitly present, this system of effective equations has the exact same form of its classical counterpart~\cite{Moradi:2013gf,Lasky:2006mg}.

For the effective model to describe a physically reasonable and well-defined collapse of a star there are a number of regularity conditions that the physical variables must satisfy. By requiring that the density at the center of the collapsing star remains well-defined at all times, we get from \cref{rho} the following condition on the mass:
\begin{equation}
\quad\quad\quad\quad\quad\quad
m(t,R) = \alpha (t) R^3 + O(R^4) , 
\quad\quad\quad\quad\quad\quad
R<<1,
\label{regularitym}
\end{equation}
where $\alpha(t)$ is a regular function of time.

Another condition comes from requiring the net force on the fluid element at the center of the star to vanish. Computing the acceleration \begin{equation}
    a^{\mu}=u^{\nu} \nabla_{\nu} u^{\mu}=\frac{1+\varepsilon}{(r^{\prime})^2} \frac{N^{\prime}}{N} \delta^{\mu}_{\, R} ,
\end{equation}
we see that a vanishing net force in $r=0$ means a vanishing $N'/N$ in $r=0$. This gives (see \cref{Nset})
\begin{equation}
\frac{2r'}{r}(\Sigma-\Pi)-\Pi ' = 0\,.
\end{equation}
In the isotropic case, the first term vanishes and this condition reduces to the vanishing of the usual force arising from the gradient of the radial pressure $\Pi '$ in $r=0$. There is however a new kind of force in the anisotropic case that does not depend on pressure gradients but on the degree of anisotropy $\Sigma - \Pi$. This force, which is outward-oriented for $\Sigma > \Pi$ and inward-oriented for $\Sigma < \Pi$, arises from gradients of the direction of anisotropy $e^\mu$ and it is there also in the Newtonian theory, it is not a relativistic effect. For this force to be regular in $r=0$, we must have $\Sigma=\Pi$ at the star's center. Actually, as we will explicitly see in the next section, $\Sigma=\Pi$ is a necessary but not sufficient condition for all the physical variables to remain well-defined in $r=0$. In fact, as the anisotropic force enters \cref{Nset} divided by $\rho + \Pi$, the regularity of the expression in $r=0$ also constrains the possible equations of state relating pressure and energy density. For some equations of state, having a regular center will not be possible.

Finally, additional conditions on the equations of state appear from the different energy conditions (weak, dominant, and strong) that we expect to be valid outside regions of Planckian curvature. Applying them to the stress-energy tensor in \cref{Tmunu} we obtain~\cite{Poisson_2004}
\begin{align}
 WEC :& \quad \rho \geq 0,  \quad \rho +\Pi\geq 0,  \quad \rho +\Sigma\geq 0 , \label{WEC} \\
 DEC :& \quad \rho -|\Pi|\geq 0, \quad  \rho -|\Sigma|\geq 0  , \label{DEC}\\
 SEC :& \quad \rho + \Pi +2\,\Sigma \geq 0 \, . \label{SEC}
\end{align}

\section{Anisotropic fluid with vanishing radial pressure} 
\label{TANG_PRES}

The set of \cref{r,mass,epsilon,rhodotset,Nset,Piset,sigma} describes a wide class of fluid collapse. Before discussing the case of a perfect fluid, where only a numerical analysis is possible, we consider the simplified but unrealistic case of vanishing radial pressure $\Pi=0$. This toy model allows us to study analytically, up to a certain degree, the interplay of pressure and quantum corrections in the semiclassical stellar collapse framework we constructed.

The first thing we need to do is equip the model with an equation of state $\Sigma=\Sigma(\rho)$. As the whole point of considering this toy model is its simplicity, not its physical accuracy, looking for a complicated EOS would be counterproductive. We thus consider the linear equation of state $\Sigma=\omega \rho$, with constant $\omega$. Assuming all energy conditions to be satisfied far away from the quantum region, the conditions in \cref{WEC,DEC,SEC} give $-1/2\le \omega \le 1$. This is the range of values of $\omega$ that we will consider in this section.

Inserting $\Pi=0$ and $\Sigma=\omega \rho$ in the effective equations of motion we get
\begin{align}
& \Dot{r}^2= N^2 \bigg( \frac{2 G m}{r} +\varepsilon\bigg)
     \,\bigg[ 1-\frac{\gamma^2 \Delta}{r^2} \left( \frac{2 G m}{r}+\varepsilon \right) \bigg]  \label{revolution} ,\\
& \Dot{m}=0 \label{massindepend}, \\
&\Dot{\varepsilon}=4\omega(1+\varepsilon) \frac{\Dot{r}}{r} , \label{varepsilononideal} \\
&\Dot{\rho}=-\rho \left(\frac{\Dot{r}'}{r'}+\frac{2 \Dot{r}}{r} \right), 
\label{rhodottang}\\
&\frac{N'}{N}=2\omega\frac{ r'}{r} .
\label{lapsenonideal}
\end{align}
In the classical limit $\Delta \rightarrow 0$ we obtain the classical EOMs~\cite{Singh:1997iy,Malafarina:2010xs} and $\omega=0$ gives the dust case. Notice that as there is no radial pressure, and so no interaction between different shells, the total gravitational mass inside each shell remains constant in time. This is represented by the fact that $m(R)$ is time-independent, as given by \cref{massindepend}. In the classical theory, $\varepsilon(t,R)$ is related to the total mechanical energy of the shell $R$, disregarding any internal energy contribution, and \cref{revolution} gives an energy balance law for each shell. The same remains true in the semiclassical theory, with quantum gravitational corrections to the energy balance law. A qualitative analysis of this equation, together with the evolution equation for $\varepsilon$ in \cref{varepsilononideal}, gives great insight into the physics of the model:

\begin{itemize}
  \item In the dust case, $\omega=0$, \cref{varepsilononideal} tells us that $\varepsilon (R)$ is a constant for each shell $R$. This is because dust has no internal energy and no interaction between different shells, so conservation of energy equals conservation of mechanical energy of each shell. In the classical theory, the gravitational potential energy contribution to $\varepsilon (R)$ decreases as the shell collapses, which means that the shell's kinetic energy increases. As there is no mechanism to stop this increase in kinetic energy, each shell will keep collapsing until it eventually crashes into the black hole singularity in $r=0$. Such a mechanism is however provided in the semiclassical theory by the quantum gravitational corrections to \cref{revolution}, which represent an effective contribution to the gravitational potential energy that increases as the shell collapses. Far away from the quantum region, for $r^3\gg G m\Delta$, these corrections are negligible and the collapse follows the classical trajectory. However, as $r$ approaches $r\sim (G m\Delta)^{1/3}$ (which is much bigger than the Planck scale $\ell_{\mathrm{Pl}}$ for an astrophysical mass $m(R)$), they become dominant, and the shell's kinetic energy starts to decrease. Eventually, all the kinetic energy is converted into gravitational potential energy, and the shell's trajectory has a turning point. This is the physical mechanism through which black hole singularities are avoided in the semiclassical theory.

  \item If $\omega \neq 0$, the mechanical energy of each shell is no longer constant (see \cref{varepsilononideal}), as their internal energy enters the energy balance as well. This opens new possibilities for the shells' trajectory already at the classical level. Let us focus on $\omega > 0$ (``regular'' matter) for the moment. In this case, \cref{varepsilononideal} gives $\Dot{\varepsilon}<0$ for $\Dot{r}<0$, which means that, as the shell collapses, its mechanical energy decreases, and so its internal energy increases. If the increase in internal energy is smaller than the decrease in gravitational potential energy, the shell acquires kinetic energy as it collapses. However, there are cases in which both gravitational potential energy and kinetic energy are turned into internal energy as the shell collapses. This second scenario can lead to a turning point in the shell's trajectory already at the classical level~\cite{Singh:1997iy}. The quantum corrections to \cref{revolution} become relevant as $r$ approaches the quantum scale and, as in the dust case, they ensure that the shell's trajectory develops a turning point (if one was not already developed by the classical dynamics).
  
  \item For $\omega < 0$ (``exotic'' matter), \cref{varepsilononideal} gives $\Dot{\varepsilon}>0$ when $\Dot{r}<0$. This means that, as the shell collapses, both gravitational potential energy and internal energy are converted into kinetic energy. A shell of exotic matter then cannot develop classical turning points during the collapse, only quantum ones.
   
\end{itemize}

This analysis gives a qualitative picture of the physics involved in the model. Let us now go into more detail. The linear EOS makes it possible to solve \cref{lapsenonideal} and compute analytically the lapse function, which takes the form $N=A(t) r^{2 \omega}$. The integration function $A(t)$ can be reabsorbed into the definition of time $t$. However, to keep the right dimensionality of $N$ we write 
\begin{equation}
    N(t,R) = \left( \frac{r(t,R)}{r_N} \right)^{2 \omega}  ,
\end{equation}
with $r_N$ being a constant with the dimension of a length. Importantly, as $r(t_0, R)=R$, already at the initial time we obtain a vanishing lapse in $R=0$ for $\omega>0$, a diverging lapse in $R=0$ for $\omega<0$, and $N(t_0,R)=1$ in the dust case $\omega=0$. As a consequence, outside the dust case, the metric is not regular at the star's center (it is degenerate for $\omega>0$ and singular for $\omega<0$) and it can be explicitly checked that the Kretschmann scalar diverges there as well. Furthermore, there is no condition on the physical variables of the model that can be imposed to avoid it. So the conditions of vanishing radial pressure and linear equation of state for the tangential pressure are incompatible with the condition of regularity of the star's center. As already noticed in~\cite{Malafarina:2010xs}, the same is true at the classical level as well. However, for our purposes, this singularity is not as bad as it may seem. In fact, since the density remains regular at the center and, as it will be shown later on, the dynamics of each shell is decoupled from the others, the formation of shell-crossing singularities in the outer region of the star is not affected by the singular dynamics of the center. This notwithstanding, to study a completely regular scenario, we will eventually consider a density profile that describes the collapse of a thick spherical shell of matter with no center. As this toy model's purpose is to study the interplay of pressure and quantum corrections in a simplified setting, and not to describe a realistic stellar collapse, this choice of density profile poses no limitations to our goal.

\Cref{varepsilononideal} can be straightforwardly solved as well to get
\begin{equation}
    \varepsilon(t,R)=-1+B(R)r^{4\omega},
\end{equation}
with $B(R) > 0$. We can give a physical meaning to $B(R)$ by noticing that at the initial time $t_0$ each shell starts with a given Newtonian ``kinetic energy'' per unit mass $K(R)~\coloneqq  ~\dot{r}(t_0,R)^2/2$. \Cref{revolution} then gives
\begin{equation}
2 K(R)=N(t_0,R)^2 \left( \frac{2 G m(R)}{R}+\varepsilon(t_0,R) \right) \left[1-\frac{\gamma^2 \Delta}{R^2} \left(\frac{2 G m(R)}{R}+\varepsilon(t_0,R)  \right) \right] ,
\label{KB}
\end{equation}
which, together with $r(t_0,R)=R$, leads to
\begin{equation} \label{eps_K}
    \varepsilon(t,R)=-1+\left[1-\frac{2 G m}{R}+\frac{R^2}{2 \gamma^2 \Delta}\left( 1-\sqrt{1- 8 K\frac{\gamma^2 \Delta \,r_N^{4\omega}}{R^{4\omega+2}}} \,\right)\right]\left(\frac{r}{R}\right)^{4 \omega} .
\end{equation}
\Cref{KB} has two different solutions for $B(R)$, but only the one chosen has the correct classical limit. Requiring the absence of trapped regions at the initial time, $r(t_0,R) = R > 2 G m (R)$ for each $R$, which allows us to study stellar collapse before any black hole is formed, guarantees that $B(R) > 0$, or equivalently that $\varepsilon(t,R) >-1$ at any time and $R$. Positivity of the square root in \cref{eps_K} gives an upper bound for the Newtonian kinetic energy
\begin{equation}
    0 \le K(R) \le K_{\mathrm{max}}(R) \coloneqq \frac{R^2}{8 \gamma^2 \Delta} \left( \frac{R}{r_N}\right)^{4\omega} .
    \label{Kmax}
\end{equation}
The existence of such a bound is a purely quantum gravitational effect and it is a result of the resolution of the black hole singularity in the semiclassical framework. In fact, if such a singularity is avoided, each shell must necessarily have a turning point ($\dot{r}=0$) at some minimum areal radius. If this is the case, however, the shell $R$ cannot have an arbitrarily high Newtonian kinetic energy $K(R)$ at some finite areal radius $r(t_0,R)$ in the past. As the value of $K_{\mathrm{max}}(R)$ is several orders of magnitude bigger than any physically reasonable value of $K(R)$ in the astrophysical collapse of a star, \cref{Kmax} poses no physical constraints on the space of initial data.

We can then study the behavior of the collapse for different initial profiles of the Newtonian kinetic energy, i.e. different initial profiles of $\Dot{r}(t_0,R)$. The function $K(R)$ is completely arbitrary provided that it is bounded by $K_{\mathrm{max}}(R) $. A convenient choice is given by $K(R)= K_\eta (R)  \coloneqq \eta K_{\mathrm{max}}(R)$, with $\eta \in [0,1]$. With this choice, $\varepsilon(t,R)$ becomes
\begin{equation} \label{eps_delta}
    \varepsilon_{\delta}(t,R)=-1+\left[1-\frac{2 G m}{R}+ \delta \frac{R^2}{2 \gamma^2 \Delta}  \right]\left(\frac{r}{R}\right)^{4 \omega} ,
\end{equation}
where $ [0,1] \ni \delta = 1-\sqrt{1-\eta} $. So we are considering a 1-parameter family of initial profiles for $\Dot{r}(0,R)$ controlled by the parameter $\delta$. A star initially at rest is given by $\delta=0$ and a star having Newtonian kinetic energy $K_{\mathrm{max}}(R) $ at the initial time is given by $\delta=1$. At a given $\delta$, the last equation of motion, the EOM for $r(t,R)$ in \cref{revolution}, is given by
\begin{equation} \label{tang_pres}
\begin{split}
 \Dot{r}^2(t,R) = &\left( \frac{r}{r_N} \right)^{4 \omega}\bigg[\frac{2 G m(R)}{r}-1+\left(1- \frac{2 G m(R)}{R} + \delta \frac{R^2}{2 \gamma^2 \Delta}\right)\left( \frac{r}{R} \right)^{4\omega} \bigg]  \\
 & \times  \bigg\{ 1-\frac{\gamma^2 \Delta}{r^2} \bigg[ \frac{2 G m(R)}{r}-1+\left(1- \frac{2 G m(R)}{R} +\delta \frac{R^2}{2 \gamma^2 \Delta}\right) \left( \frac{r}{R} \right)^{4\omega} \bigg] \bigg\}  \, .
\end{split}
\end{equation}
From this equation, it is clear that the shells are decoupled from each other and that they undergo their own independent evolution. This means that we can numerically solve \cref{tang_pres} to find the trajectory of each shell $R$ separately, and then check if these trajectories cross to form shell-crossing singularities. Notice however that although \cref{tang_pres} can be used to formally continue the evolution of the shells past a SCS, the complete set of EOMs (\cref{revolution,massindepend,varepsilononideal,rhodottang,lapsenonideal}) is no longer well defined there and the physics of the SCS needs to be explicitly taken into account to study the future evolution of the system.


Before plunging into the numerics, there are still some qualitative aspects of the dynamics expressed by \cref{tang_pres} that can be investigated analytically. Turning, or bouncing, points of the shells' trajectories play an important role in shell-crossing formation. From \cref{tang_pres} we see that there are two possible zeroes of $\dot{r}$: the one coming from the vanishing of the expression inside square brackets, and the one coming from the vanishing of the expression inside curly brackets. The vanishing of the expression inside square brackets gives 
\begin{equation}
\left(1- \frac{2 G m}{R} + \delta \frac{R^2}{2 \gamma^2 \Delta} \right) \left( \frac{r}{R} \right)^{4\omega} r-r+2Gm=0 ~.
\label{bouncec}
\end{equation}
This is a polynomial equation in $r$ whose degree depends on $\omega$. Although a general analytic solution $r(\omega,R)$ cannot be found, we get an interesting lower bound for $r$ by rewriting \cref{{bouncec}} as
\begin{equation}
r-2Gm = \left(1- \frac{2 G m}{R} + \delta \frac{R^2}{2 \gamma^2 \Delta}\right)\left( \frac{r}{R} \right)^{4\omega} r >0 ~.
\label{bounceclb}
\end{equation}
This tells us that this turning point can only take place for $r(t,R)>2Gm(R)$, namely before the trajectory ever enters a trapping horizon. Given this property and the fact that these turning points are the only ones present in the classical theory, as the expression in curly brackets in \cref{tang_pres} goes to 1 in this limit, we will refer to the solutions $r_c (\omega,R)$ of \cref{bouncec} as the \emph{classical turning points} or the \emph{classical bounce points}.

The vanishing of the expression inside curly brackets gives
\begin{equation}
    \left(1- \frac{2 G m}{R} + \delta \frac{R^2}{2 \gamma^2 \Delta}\right)\left( \frac{r}{R} \right)^{4\omega} r=\frac{r^3}{\gamma^2 \Delta} +r-2Gm ~ .
    \label{bounceq}
\end{equation}
The solutions $r_q (\omega,R)$ to this equation will be referred to as the \emph{quantum turning points} or \emph{quantum bounce points}. As the left-hand side of the equation is explicitly positive, we get a lower bound $r>r_{\mathrm{lb}}$ also for the quantum turning points, with 
\begin{equation}
    r_{\mathrm{lb}}=\frac{\left( G m\gamma^2 \Delta + \sqrt{\left(\gamma^2\Delta/3\right)^3+ \left(G m\gamma^2 \Delta\right)^2}\right)^{2/3} -  \left(\gamma^2 \Delta/3 \right)}{\left( G m\gamma^2 \Delta + \sqrt{\left(\gamma^2\Delta/3\right)^3+ \left(G m\gamma^2 \Delta\right)^2}\right)^{1/3}} \, ~. 
    \label{lb}
\end{equation} 
Expanding this expression in powers of $\Delta/ \left[Gm(R)\right]^2$ we get
\begin{equation}
r_{\mathrm{lb}}=Gm \left\{ \left[ 2\gamma^2 \frac{\Delta}{(Gm)^2}\right]^{1/3}-\frac{1}{3}\left[\frac{\gamma^2 \Delta}{\sqrt{2}(Gm)^2} \right]^{2/3} +O\left[\frac{\Delta}{(G m)^2}\right] \right\}\, ,
\label{lbexp}
\end{equation}
which, at order $\left[\Delta/(Gm)^2\right]^{1/3}$, turns out to be equal to the analogous lower bound found in the marginally bound dust case~\cite{Fazzini:2023ova,Giesel:2023hys}. This suggests that the presence of a non-vanishing tangential pressure does not modify significantly the effect of quantum gravity repulsion. Notice however that the expansion in \cref{lbexp} does not hold for shells with 
$Gm\lesssim\sqrt{\Delta} \,$. Furthermore, for $\omega=0$, we get
\begin{equation}
r_q (\omega=0, R)=\left(2 G m \gamma^2 \Delta \right)^{1/3}\left\{1 + \frac{\varepsilon_0}{6 G m}\left(2 G m \gamma^2 \Delta \right)^{1/3} + O\left[\frac{\Delta}{(Gm)^2}\right]\right\}, 
\label{rqdust}
\end{equation}
\begin{equation}
\varepsilon_0=-\frac{2 G m}{R} + \delta \frac{R^2}{2 \gamma^2 \Delta},
\end{equation}
which is the same value found in~\cite{Cafaro:2024vrw} for the quantum bounce radius in the bound and unbound dust case, and it holds for any shell of each initial dust distribution~\cite{Cipriani:2024nhx}. It is interesting to notice that the bigger the mechanical energy $\varepsilon_0$, which is a constant of motion in the dust case, the further from the center the shell will bounce. While counterintuitive, this is a simple consequence of the quantum corrections in \cref{revolution}, as they provide a stronger repulsion for bigger mechanical energy $\varepsilon$.

\begin{figure} 
    \centering
    \captionsetup{justification=raggedright}
     \begin{subfigure}{0.45\textwidth}
         \centering
         \includegraphics[width=\textwidth]{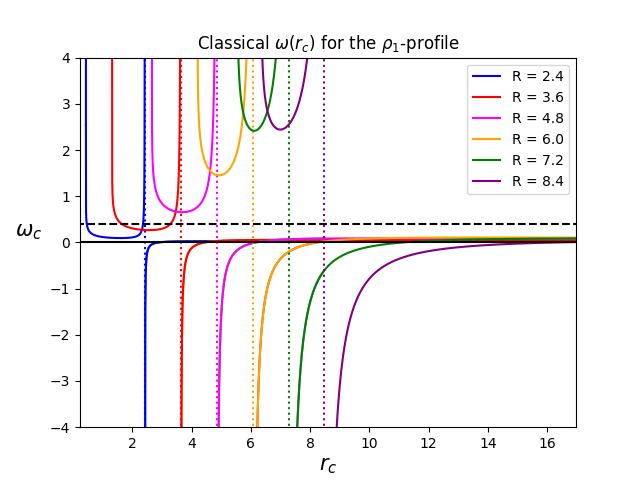}
         \caption{}
         \label{fig:w_c}
     \end{subfigure}
     \begin{subfigure}{0.46\textwidth}
         \centering
         \includegraphics[width=\textwidth]{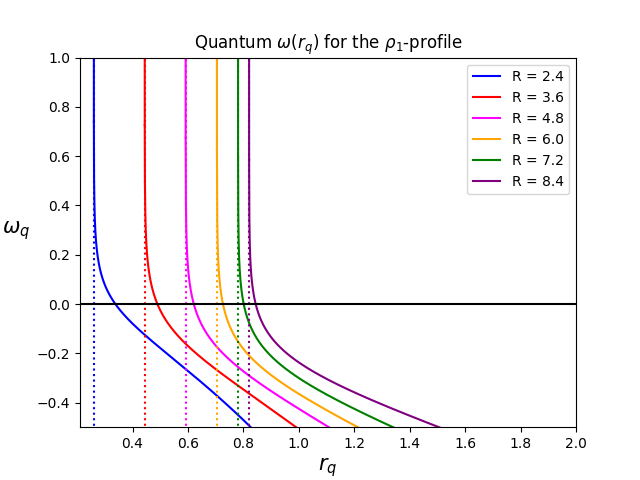}
         \caption{}
         \label{fig:w_q}
     \end{subfigure}
     \hspace{3em}%
     \caption{Plot of the functions $\omega_c (r_c,R)$ (left) and $\omega_q (r_q,R)$ (right) for six different values of $R$. The colored vertical dotted lines represent $r_c=R$ (left) and $r_q = r_{\mathrm{lb}}$ (right). The black dashed horizontal line in the left plot represents $\omega = \omega_* := 0.4$. Natural units and constants are set to $c=G=k_{\mathrm{B}}=1$, $\gamma^2\Delta=0.1$, $\delta=0.001$, $R_B=10$, and $\rho_0=4\times 10^{-3}$.}
     \label{fig:w}
\end{figure}

It is difficult to solve \cref{bouncec,bounceq} for the bounce radii $r_c (\omega,R)$ and $r_q (\omega,R)$ given a shell $R$ and a EOS parameter $\omega$. Interestingly, it is instead straightforward to solve them for the value of $\omega(r,R)$ such that the bounce of the shell $R$ takes place at the areal radius $r$:
\begin{align}
& \omega_c (r_c ,R)=\frac{1}{4}\left[\frac{\ln{ (1-\frac{2Gm}{r_c})}- \ln{ (1-\frac{2Gm}{R} + \delta \frac{R^2}{2 \gamma^2 \Delta})}}{\ln{(\frac{r_c}{R})}} \right] \label{bounce_c} , \\
& \omega_q (r_q ,R) =\frac{1}{4}\left[\frac{\ln{ (1-\frac{2Gm}{r_q} + \frac{r_q^2}{\gamma^2 \Delta})}-\ln{ (1-\frac{2Gm}{R}+ \delta \frac{R^2}{2 \gamma^2 \Delta})}}{\ln{(\frac{r_q}{R})}}\right]   \label{bounce_q}   .
\end{align}
Notice that, thanks to the chosen initial conditions and the lower bounds just found, all the arguments of the logarithms remain positive throughout the shells' evolution. In order to plot these functions, we need to specify an initial density profile for the star. We consider for this purpose the profile
\begin{align}
\rho_1(t_0,R) &=
\begin{cases}
   \frac{\rho_0}{2} \big(1+\cos(\pi R/R_B)\big) & \text{if} \quad R < R_B,  \\
    0   & \text{if} \quad R \ge R_B  ,
\end{cases}
\label{density_profile_1} 
\end{align}
%
%
where $\rho_0$ and $R_B$ are respectively the energy density in $R=0$ and the comoving radial coordinate of the boundary of the star. Given this density profile, the mass profile $m(R)$ can be obtained using \cref{mass_def}. The plots of the functions $\omega_c (r_c,R)$ and $\omega_q (r_q,R)$ for several fixed values of $R$ are given in Fig.~\ref{fig:w}.

Let us start the discussion from $\omega_c (r_c ,R)$. For each $R$, this function is positive for $2Gm(R)<r_c<R$, with vertical asymptotes in $r_c=2Gm(R)$ and $r_c=R$. The function does not exist in $r_c\leq2Gm(R)$ and $r_c=R$ as there is no $\omega$ such that \cref{bouncec} is satisfied there for arbitrary $\delta$. The physical scenario we are interested in is one where $\omega$ of the EOS is fixed for the entire star, i.e. for all the shells, and we want to study when each shell bounces given the chosen $\omega$. So, fixing $\omega=\omega_*$ (taken to be $\omega_*=0.4$ in the plot in Fig.~\ref{fig:w_c}), the shell $R$ will bounce at the areal radius $r_c$ such that $\omega_c (r_c ,R)=\omega_*$, namely at an intersection between the two curves $\omega=\omega_c (r_c ,R)$ and $\omega=\omega_*$. As we are considering initially collapsing configurations for which $r(t,R)<R$, and $\omega_c (r_c ,R)>0$ for $r_c<R$, there cannot be any classical turning points for $\omega\leq0$. However, there will be classical turning points for $\omega>0$. A classical bounce occurs for all the shells for which 
\begin{equation}
    \displaystyle\min_{r_c<R}\big\{\omega_c(r_c,R)\big\} \leq \omega_* ,
\end{equation} 
with the bouncing radius $r_c$ given by the rightmost intersection between $\omega=\omega_c (r_c ,R)$ and $\omega=\omega_*$. For the density profile and initial conditions chosen in Fig.~\ref{fig:w_c}, the interior of the star bounces classically while the exterior does not. This will inevitably form a shell-crossing singularity already at the classical level. We found numerically that this is the case for generic density profiles and initial conditions (with the star starting the collapse outside any trapped region) for $\omega>0$ in this toy model.

From Fig.~\ref{fig:w_q} it is easy to see that there is always an intersection between $\omega=\omega_q (r_q ,R)$ and $\omega=\omega_*$ for $-1/2\leq\omega\leq 1$, which means that every shell $R$ of the star will have a quantum turning point. From this analysis, we cannot infer whether a shell-crossing singularity will develop due to the quantum bounce. In the Oppenheimer-Snyder collapse the quantum bounce of the shells is synchronized in such a way that no SCS develops~\cite{Fazzini:2023scu}. We will however see that they do generally develop in this case.

These results are consistent with the qualitative analysis of the energy balance law we carried out at the start of the section and they offer a more detailed account of the physics of the model. Before moving on, let us point out a peculiar feature of the shells' dynamics that can be inferred from Fig.~\ref{fig:w}. For $\omega < 0$, no classical bounce can happen. So shells will necessarily undergo a quantum bounce, after which they will expand. At this point, however, the shell can get to $r(t,R)>R$ where $\omega=\omega_c (r_c ,R)$ and $\omega=\omega_*$ have an intersection. This means that the shell will have a classical turning point, producing a further contraction, which will be followed by yet another quantum bounce and so on. These infinitely bouncing trajectories will be shown shortly in the numerical simulations of stellar collapse in this toy model.

\begin{figure}
    \centering
    \captionsetup{justification=raggedright}
    \hfill
     \begin{subfigure}{0.35\textwidth}
         \centering
         \includegraphics[width=\textwidth]{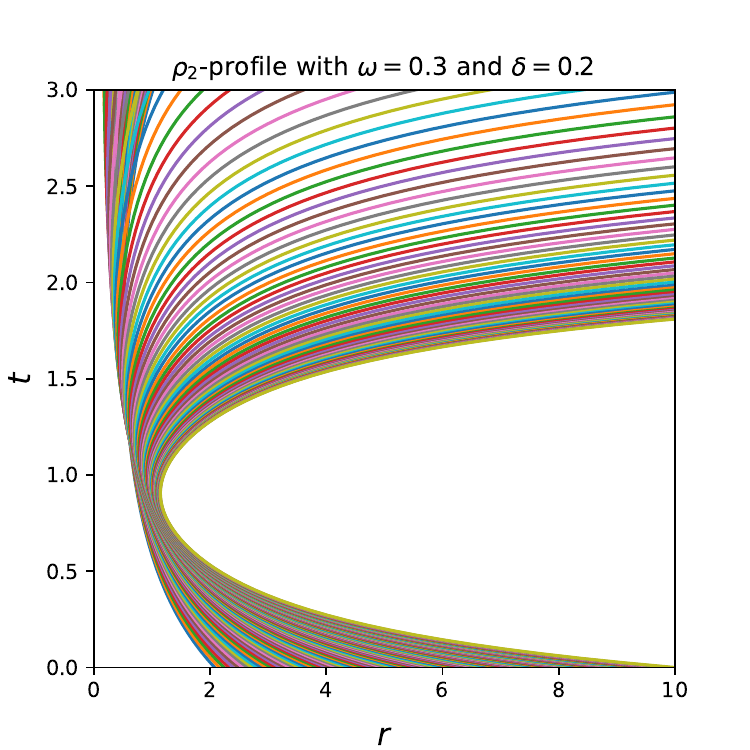}
         \caption{}
         \label{fig:cos2_0_3_D1}
     \end{subfigure}
     \hfill
     \begin{subfigure}{0.35\textwidth}
         \centering
         \includegraphics[width=\textwidth]{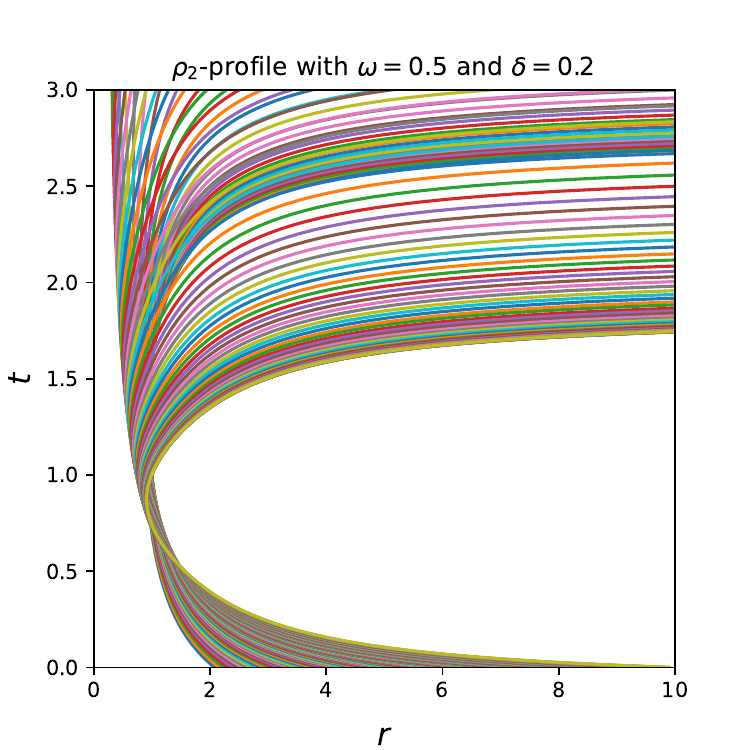}
         \caption{}
         \label{fig:cos2_0_5_D1}
     \end{subfigure}
     \hfill
     \hspace{3em}%
     \caption{Shells dynamics and shell-crossing singularity formation for $\rho_2$ in \cref{density_profile_2} with $\omega=0.3$ (left)  and $\omega=0.5$ (right). Natural units and constants are set to $c=G=k_{\mathrm{B}}=1$, $\gamma^2\Delta=0.1$, $r_N=1$, $\delta=0.2$, $\rho_0 =2 \times 10^{-3}$, $a=4$, and $R_B=6$.}
     \label{fig:cos2_D1}
\end{figure}

\begin{figure} 
    \centering
    \captionsetup{justification=raggedright}
     \begin{subfigure}{0.32\textwidth}
         \centering
         \includegraphics[width=\textwidth]{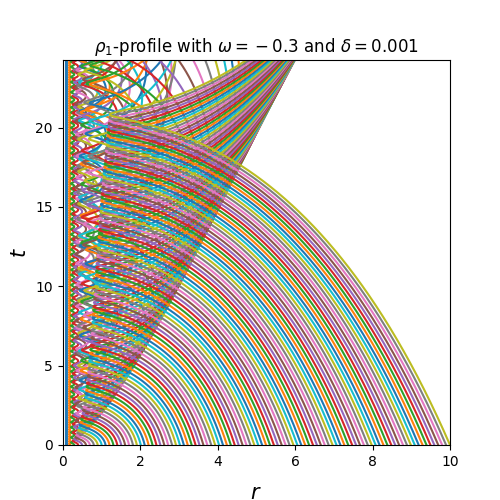}
         \caption{}
         \label{fig:COS_-0_3}
     \end{subfigure}
     \begin{subfigure}{0.32\textwidth}
         \centering
         \includegraphics[width=\textwidth]{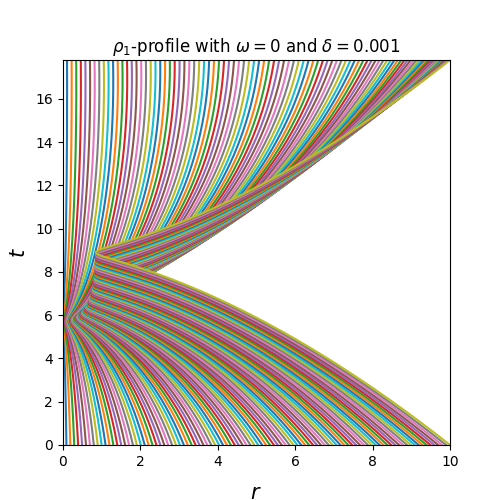}
         \caption{}
         \label{fig:COS_0}
     \end{subfigure}
     \begin{subfigure}{0.32\textwidth}
         \centering
         \includegraphics[width=\textwidth]{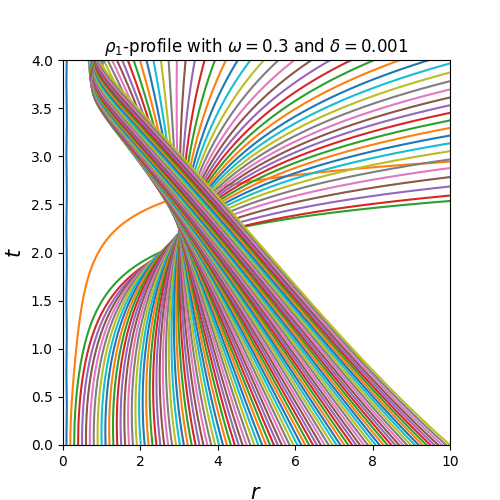}
         \caption{}
         \label{fig:COS_0_3}
     \end{subfigure}
     \hspace{3em}%
     \caption{Shells dynamics and shell-crossing singularity formation for $\rho_1$ in \cref{density_profile_1} with $\omega=-0.3$ (left),  $\omega=0$ (center), and $\omega=0.3$ (right). Natural units and constants are set to $c=G=k_{\mathrm{B}}=1$, $\gamma^2\Delta=0.1$, $r_N=1$, $\delta=0.001$, $\rho_0 =4 \times 10^{-3}$, and $R_B=10$.}
     \label{fig:COS}
\end{figure}

We are ready to numerically integrate the EOM in \cref{tang_pres}. Up until now, the singularity in the center of the star did not enter our analysis. However, as soon as we consider the acceleration $\ddot{r}(t_0,R)$ of the shells at the initial time
\begin{equation}
\begin{split}
\ddot{r}(t_0,R)= &\frac{\omega }{2 \gamma^2 \Delta r_N^{4\omega}} R^{4\omega-1}\left[1- (\delta-1)^2\right] \\
 & + \frac{ R^{4\omega-1}}{2 r_N}\left[\left( -\frac{2m(R)}{R} + 4\omega B(R)\right)(1-\delta) + \delta^2 \frac{R^2}{2 \gamma^2 \Delta}\right] ,
\end{split}
\label{accelerationeq}
\end{equation}
we see that it diverges as $R\rightarrow 0$ (where \cref{regularitym} holds) for any given $\delta$ and $-1/2<\omega<1/4$ ($\omega\neq0$). The dynamics of the star's center is thus singular, as already anticipated. Nevertheless, as an arbitrary density (and consequently mass) profile remains regular at the center and the dynamics of each shell is decoupled from the others (see \cref{tang_pres}), the formation of shell-crossing singularities in the outer region of the star is not affected by the singular dynamics of the center. For this reason, together with a completely regular density profile $\rho_2$ that describes the collapse of a thick spherical shell of matter with no center, we will also numerically integrate the EOMs for the profile $\rho_1$ in \cref{density_profile_1}. We take the regular profile $\rho_2$ to be given by
\begin{align}
\rho_2(t_0,R) &=
\begin{cases}
   \rho_0 \cos^2\big(\frac{\pi}{2a} (R-R_B)\big) & \text{if} \quad |R-R_B| < a,  \\
    0   & \text{otherwise}   .
\end{cases}
\label{density_profile_2} 
\end{align}
\begin{figure}
    \centering
    \captionsetup{justification=raggedright}
     \begin{subfigure}{0.32\textwidth}
         \centering
         \includegraphics[width=\textwidth]{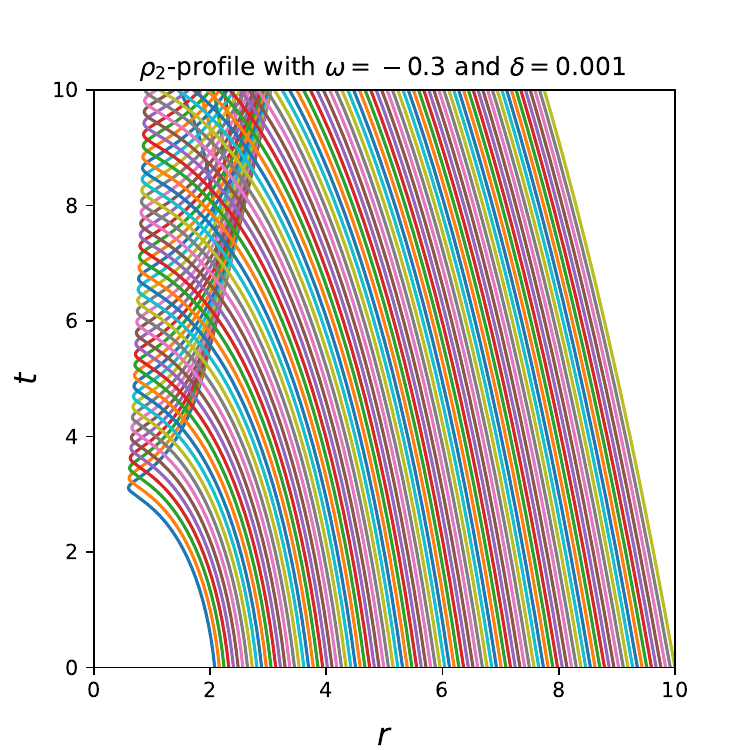}
         \caption{}
         \label{fig:cos2_-0_3}
     \end{subfigure}
     \begin{subfigure}{0.32\textwidth}
         \centering
         \includegraphics[width=\textwidth]{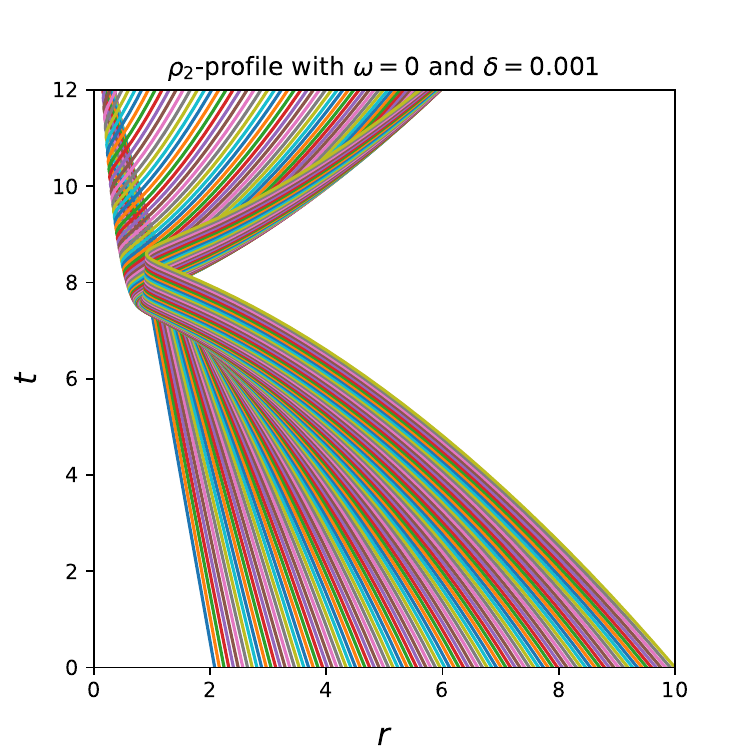}
         \caption{}
         \label{fig:cos2_0}
     \end{subfigure}
     \begin{subfigure}{0.32\textwidth}
         \centering
         \includegraphics[width=\textwidth]{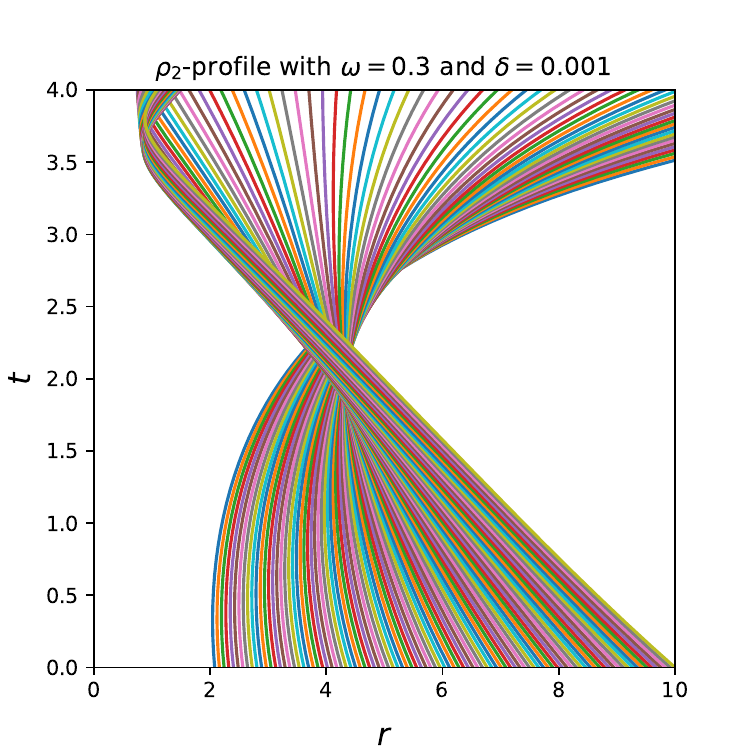}
         \caption{}
         \label{fig:cos2_0_3}
     \end{subfigure}
     \hspace{3em}%
     \caption{Shells dynamics and shell-crossing singularity formation for $\rho_2$ in \cref{density_profile_2} with $\omega=-0.3$ (left),  $\omega=0$ (center), and $\omega=0.3$ (right). Natural units and constants are set to $c=G=k_{\mathrm{B}}=1$, $\gamma^2\Delta=0.1$, $r_N=1$, $\delta=0.001$,  $\rho_0 =2 \times 10^{-3}$, $a=4$, and $R_B=6$.}
     \label{fig:cos2}
\end{figure}
Finally, using a Runge-Kutta 4 algorithm it is possible to numerically integrate \cref{tang_pres} to obtain the dynamics shown in Fig.~\ref{fig:cos2}, Fig.~\ref{fig:cos2_D1}, and Fig.~\ref{fig:COS}. These figures show the results of the numerical integrations for $\rho_1$ and $\rho_2$ for different values of $\omega$ and $\delta$. Interestingly, the value of $\delta$ cannot be taken to be vanishing if we want all the collapsing scenarios considered to have a well-defined initial evolution. In fact, studying \cref{accelerationeq} for $\rho_1$ when $\omega > 0$ and $\delta=0$, it is straightforward to see that the initial acceleration of the shells is positive near the center of the star (where $R<<1$ and \cref{regularitym} holds) and negative in the outer regions, which means that shells will inevitably cross an instant after the initial time. So we must have $\delta>0$ to avoid this situation.

The first thing to notice from these plots is that shell-crossing singularities form in all scenarios considered. This feature is unaltered if one considers different initial profiles and initial conditions. Let us start our discussion from the regular case of the initial profile $\rho_2$ in Fig.~\ref{fig:cos2}. For $\omega=0$, namely the dust case, a shell-crossing singularity develops before the quantum bounce. This behavior is a peculiar feature of density profiles with two tails, such as $\rho_2$. This is consistent with previous investigations of the dust case~\cite{Fazzini:2023ova,Giesel:2023hys,Cipriani:2024nhx}. Even if the independence of the shells' evolution allows us to integrate the system in the future of this event, where we can see several more SCS happening, the actual physical evolution of the system will depend on the physics of the first shell-crossing. For $\omega<0$, a SCS develops just after the quantum bounce of the most interior shell and we see no classical SCS. For $\omega>0$, however, a SCS singularity develops much before the quantum region is reached. In particular, if $\omega$ (and so the pressure) is sufficiently big and $\delta$ (and so the initial kinetic energy) sufficiently small, then the increase in internal energy as the shells collapse leads to a very brief phase of contraction followed by a classical bounce. This is exactly what happens to the inner shells in Fig.~\ref{fig:cos2_0_3}. On the contrary, the decrease in gravitational energy wins over the increase in internal energy for the outer shells, and so they keep collapsing until the quantum bounce. This leads to the peculiar dynamics shown in Fig.~\ref{fig:cos2_0_3}, which was already inferred also from Fig.~\ref{fig:w_c}, and the formation of a shell-crossing singularity.

As the quantum corrections present in our semiclassical model are negligible in the classical region where SCS generally develop for $\omega>0$, these are not the events we want to study. We want to investigate if the interplay of pressure and quantum corrections is able to avoid the formation of SCS where the quantum corrections are relevant. To focus on these events, in this toy model, we could either decrease the value of $\omega$, consider only shells that are further away from the center, or increase the value of $\delta$. For the forthcoming discussion, we choose the latter. In fact, for a sufficiently big initial kinetic energy (close to the value of $K_{\mathrm{max}}$ in \cref{Kmax}), there is not enough time to convert all the kinetic energy in internal energy before the quantum region is reached, and so the classical bounce is avoided. This is exactly what happens in the scenarios shown in Fig.~\ref{fig:cos2_D1}, where $\delta=0.2$. Even in these cases the shells' evolution leads to a shell-crossing singularity, but this time in the quantum region of spacetime.

All this shows that the presence of tangential pressure does not prevent the formation of shell-crossing singularities. As one would naively think that radial pressure is more important in preventing shells from crossing than tangential pressure, this result is interesting but not altogether surprising. We will see in the next section that, perhaps more unexpectedly, even the inclusion of radial pressure is not able to prevent the formation of shell-crossing singularities.

Let us now briefly comment also the plots for the initial profile $\rho_1$ in Fig.~\ref{fig:COS}. As can be seen from the plots, the behavior of the exterior shells of the star is completely analogous to the behavior we have seen for the spherical matter shell with no center. This confirms that the outer region of the star is not affected by the singular dynamics of the center. From the central plot for $\omega=0$ (where there is no singularity), we see that the center of the star undergoes a quantum bounce without forming shell-crossings. A SCS singularity is however formed just after the bounce as we move towards the exterior. Fig.~\ref{fig:COS_-0_3} exhibits the phenomenon of infinitely bouncing trajectories for $\omega<0$ that was inferred from the analysis of turning points. Notice that all shells present this behavior for $\omega<0$, also the ones in Fig.~\ref{fig:cos2_-0_3}. It is only more apparent as we get closer to the center of the profile $\rho_1$ because the acceleration of the shells explodes there, making the frequency of the oscillations much higher. At the same time, however, as we move closer and closer to the center, we also see that the amplitude of these oscillations is damped out. This is because both values of the turning points $r_c$ and $r_q$ between which the trajectories oscillate tend to zero as $R\rightarrow0$. This is also the reason why the center $R=0$ of the star remains at $r(t,0)=0$ even if its acceleration is infinite. As any shell $R>0$ has $r_{\mathrm{lb}}>0$, no shell $R>0$ can come in contact with the singular center $R=0$. This is one more reason why the singularity does not interfere with the overall evolution of the collapse. For $\omega>0$, a SCS singularity is formed already at the classical level. Although the interior shells move towards the exterior, the singular center $R=0$ remains in $r(t,0)=0$.

\section{Perfect fluid}
\label{sec:perfect}

The assumption of vanishing radial pressure allowed us to analytically study the interplay of pressure and quantum corrections in a simplified setting. It also helped us to develop intuition for the physics of the semiclassical stellar collapse framework we constructed. However, the absence of radial pressure is far from being realistic. A more realistic scenario is found by treating the star as a perfect fluid with $\Pi=\Sigma \eqcolon p$. The stress-energy tensor will then take the usual form $T^{\mu}_{\; \nu}=\mathrm{diag}(-\rho, p, p, p)$, which is straightforwardly recovered from the general case in \cref{Tmunu} by enforcing the equality of radial and tangential pressure. The effective EOMs for the perfect fluid case read (see \cref{r,mass,epsilon,rhodotset,Nset,Piset,sigma})
\begin{align}
&\Dot{r}^2= \, N^2 \bigg( \frac{2 G m}{r} +\varepsilon\bigg)
     \,\bigg[ 1-\frac{\gamma^2 \Delta}{r^2} \left( \frac{2 G m}{r}+\varepsilon \right) \bigg],  \label{ciccio0}  \\
&\Dot{m}=-4\pi p\, r^2 \Dot{r} \label{mass_per} , \\
&\Dot{\varepsilon}=-2(1+\varepsilon)\frac{p'}{\rho+p}\frac{\Dot{r}}{r'} \label{epsilon_per} ,\\
& \Dot{\rho}=-p^{\prime}\frac{\Dot{r}}{r'}-(\rho+p)\left( \frac{\Dot{r}'}{r'}+2\frac{\Dot{r}}{r}\right)   \label{density_per}, \\
& \frac{N'}{N}=-\frac{p^{\prime}}{\rho+p} .
\label{NNN}
\end{align}
Differently from the vanishing radial pressure case, the total mass-energy $m(t,R)$ contained inside the shell $R$ is no longer a constant of motion. In fact, its time dependence satisfies \cref{mass_per}, which is the same equation it also satisfies classically. The right-hand side of \cref{mass_per} is the volume-work made on the portion of the star inside the shell $R$ as it evolves. For a collapsing star ($\Dot{r}<0$) with positive radial pressure, this work is positive and $m(t,R)$ increases in time. The opposite happens for negative radial pressure, in this case the work is negative and $m(t,R)$ decreases in time. Notice however that, since $p=0$ on the outermost physical shell of the star, $\Dot{m}=0$ on the boundary of the star and so the total gravitational mass is conserved. Unfortunately, given the complexity of the EOMs, this system has to be faced entirely numerically.

The freedom provided by the choice of the equation of state $p=p(\rho)$ allows us to model the collapse of different stars like neutron stars, quark stars, red giants, etc. In this work, however, we will limit ourselves to a linear EOS $p=\omega \rho$, with $\omega= \mathrm{const}$. The numerical code can be easily adapted to other equations of state, such as polytropics $p=k \rho^\gamma$ or even more exotic ones, but we leave the detailed investigation of these scenarios for future work. Assuming all energy conditions to be satisfied far away from the quantum region, the conditions in \cref{WEC,DEC,SEC} give $-1/3\le \omega \le 1$. This is the range of values of $\omega$ that we will consider in this section.

The system of equations can be slightly simplified by noticing the existence of a conserved quantity $Q(R)$, whose derivation and properties are discussed in Appendix~\ref{conservedquantity}, given by
\begin{equation}
Q(R)=\frac{4\pi\rho^{\frac{1}{1+\omega}} r^2 r'}{\sqrt{1+\varepsilon}}\, .
\end{equation}
This quantity can be used to write
\begin{align}
&\Dot{\varepsilon}=-8\pi \omega r^2 \Dot{r}\frac{\sqrt{1+\varepsilon}}{Q(R)}\partial_R \left(\rho^{\frac{1}{1+\omega}} \right),
\label{epsilon_per'}\\
& \rho=\left(\frac{Q(R)\sqrt{1+\varepsilon}}{4\pi r^2 r'} \right)^{1+\omega}, \label{density_per'} 
\end{align}
which appreciably simplifies the numerical integration of the system. Numerical solutions of the system of \cref{ciccio0,mass_per,NNN,epsilon_per',density_per'} are obtained using the method of lines. In this method, the spatial dimension is discretized, and spatial derivatives are computed using second-order finite differences, while the time evolution is managed with a fourth-order Runge-Kutta scheme. The spatial and temporal discretization steps are chosen to meet the Courant-Friedrich-Lewy condition, $\delta_t < \delta_x $~\cite{Leveque:2002} (here the maximum allowed velocity of the fluid shells is $c=1$).


In what follows, we choose our initial configuration to be in the bound case, i.e. $\varepsilon(t_0, R) < 0$, by adopting the smooth decreasing function
\begin{equation}
    \varepsilon(t_0, R) = \varepsilon_* \frac{\exp\left(- \frac{R^2}{2\sigma^2} \right) - 1}{\sqrt{2 \pi \sigma^2}} \, ,
\end{equation}
with $\varepsilon_*$ and $\sigma$ constants such that $\varepsilon_* < \sqrt{2 \pi \sigma^2}$, so to make sure that $\varepsilon >-1$ everywhere. Given an initial density profile and the equation of state, \cref{NNN} calculated at the initial time $t_0$ can be integrated to obtain
\begin{equation}
    N(t_0, R) = \exp \left( - \int_{R_B}^R \left(\frac{p'}{p + \rho}\right) \Big|_{t_0} \dd \tilde{R} \right) {\sqrt{1+\varepsilon(t_0,R_B)}}~,
    \label{Nsimulation}
\end{equation}
where $R_B$ is the boundary shell of the star. The integration constant is selected so that $N(t_0, R_B) = \sqrt{1+\varepsilon(t_0,R_B)}$, in agreement with the discussion of the vacuum solution in \cref{sec:vacuum}. In fact, while in the classical case the lapse can be set to $N(t_0, R_B)=1$ on the boundary of the star, at the effective level Birkhoff theorem breaks down~\cite{Cafaro:2024vrw} and each collapsing star generates its own particular exterior.

As explicitly shown in the anisotropic context of the last section, in stellar collapse scenarios with pressure shell-crossing singularities can generally develop in the classical region of spacetime, even before any black hole is formed. As the quantum corrections present in our semiclassical model are negligible there, these are not the events we want to study. We want to investigate if the interplay of pressure and quantum corrections is able to avoid the formation of SCS where the quantum corrections are relevant. To focus on these events, in this section we consider initial density profiles where a black hole has already formed and the star is collapsing towards the quantum region of spacetime.

\begin{figure}
    \centering
    \captionsetup{justification=raggedright}
    \includegraphics[width=0.9\textwidth]{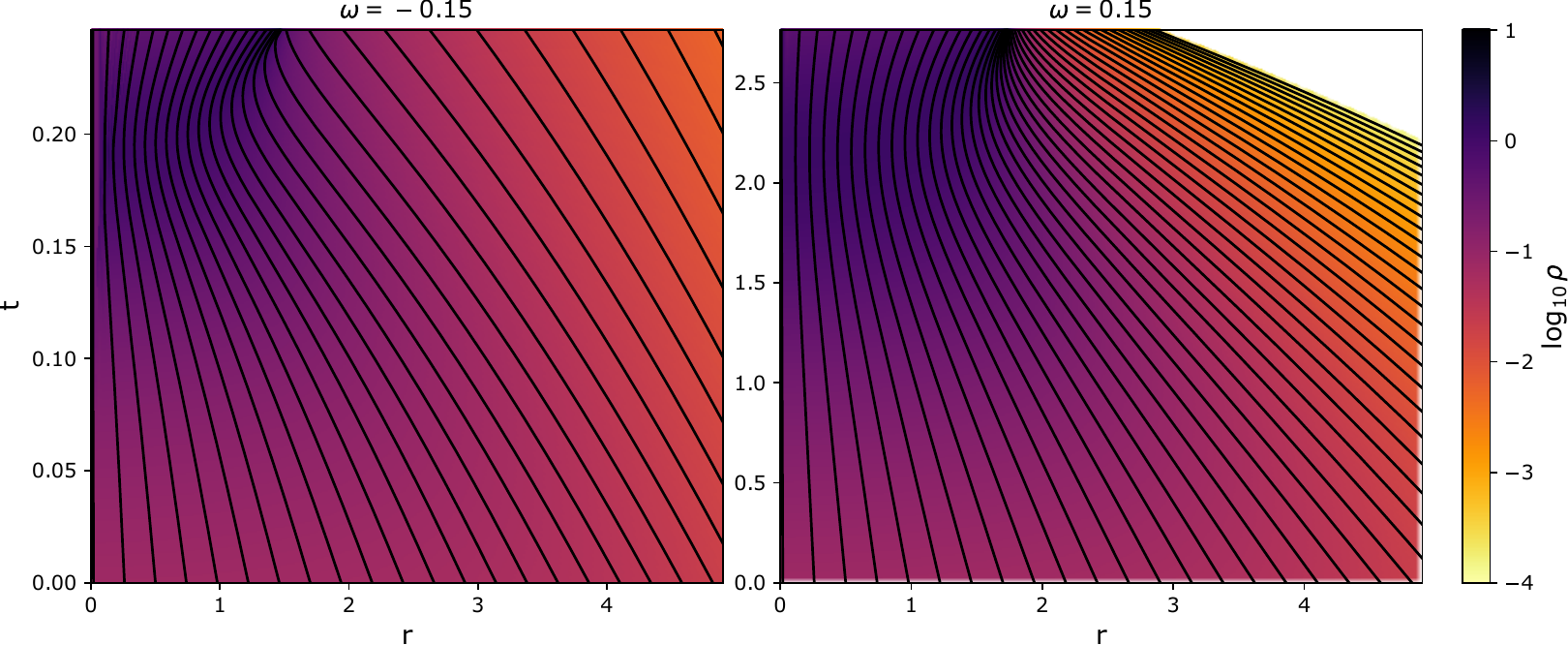}
    \caption{Shells dynamics and shell-crossing singularity formation for $\rho_3$ in \cref{rho3} with $\omega=-0.15$ (left) and $\omega=0.15$ (right). Natural units and constants are set to $c=G=k_{\mathrm{B}}=1$, $\gamma^2\Delta=0.1$, $\sigma=3.5$, $\varepsilon_*=1$, $M=30$, $R_0=11$, $\lambda=3.5$.}
    \label{tanhplot}
\end{figure}

With this setup in place, the system is fully determined and ready for numerical evolution. We will now show the numerical results of the integration of the system for several initial configurations. The first one we consider is
\begin{align}
 \rho_3 (t_0,R)= C_3 \left[1-\tanh\left( \frac{R-R_0}{\lambda}\right)\right] , 
 \label{rho3}
 \end{align}
 \begin{align}
  C_3 =\frac{M}{{\int_{0}^{R_B} 4 \pi \Tilde{R}^2 \left[1-\tanh\left( \frac{\tilde{R}-R_0}{\lambda}\right)\right] \dd \Tilde{R}}}\,,
  \label{c3}
\end{align}
with $M$ the total gravitational mass of the star. Notice that even if this profile, and all the profiles we consider later on, are analytically of non-compact support, the numerical code approximates the residual tail to zero at a finite distance from the center, making them effectively bounded. In particular, we have set the numerical boundary $R_B$ such that (see \cref{c3})
\begin{equation}
     4 \pi \int_0^{+\infty} \rho (t_0, \Tilde{R}) \Tilde{R}^2  \,\dd \Tilde{R} \: - \: M = 10^{-3} .
\end{equation}
There is no appreciable change if this difference is taken to be smaller. Simulations corresponding to positive and negative $\omega$ are shown in Fig.~\ref{tanhplot}.
These plots show with a color scale the logarithmic density $\log_{10} \rho$ as a function of the physical areal radius $r(t,R)$ and $t$. The trajectories of a few evenly-spaced shells $R$ are shown in black. Shell-crossing singularities develop just after the quantum bounce of the star in both cases, and they can be easily recognized in the plot by the converging of neighboring shells and the black color of the density's color scale. Differently from the dust and vanishing radial pressure cases, where the EOMs decouple and the dynamics of each shell can be evolved past a shell-crossing singularity, here the system is coupled and the evolution breaks down, and rightly so, as soon as a shell-crossing singularity develops. Also, it is interesting to notice that the shells that cross each other belong to the ``tail'' of the initial density profile, while the inner shells bounce regularly without crossing. This behavior was already noticed in the dust case~\cite{Fazzini:2023ova,Husain:2022gwp, Cipriani:2024nhx}, as also shown in Fig.~\ref{fig:COS_0}. Since there are no shell crossings in the homogeneous case (Oppenheimer-Snyder)~\cite{Fazzini:2023scu}, this phenomenon signals that it is the inhomogeneity of the profile that favors the shells to cross.


\begin{figure}[b]
    \centering
    \captionsetup{justification=raggedright}
    \includegraphics[width=0.9\textwidth]{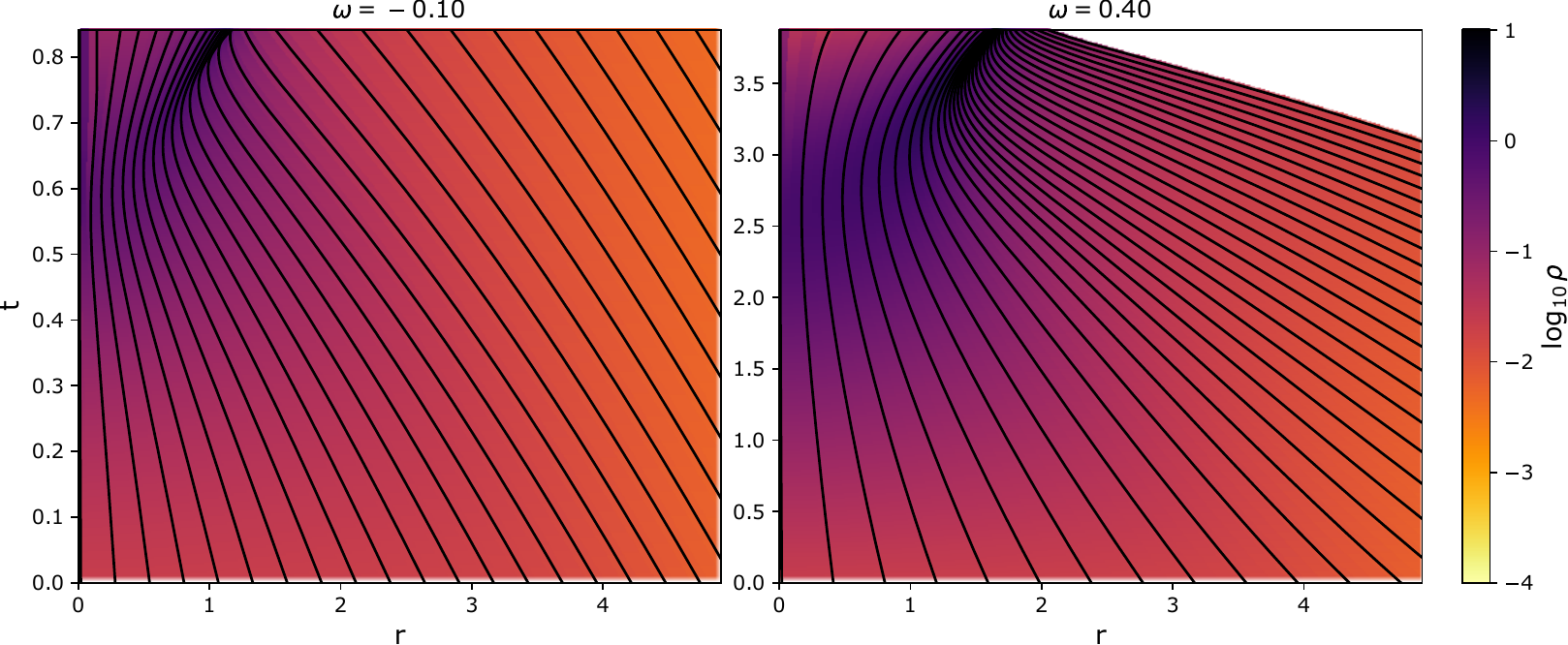}
    \caption{Shells dynamics and shell-crossing singularity formation for $\rho_4$ in \cref{rho4} with $\omega=-0.1$ (left) and $\omega=0.4$ (right). Natural units and constants are set to $c=G=k_{\mathrm{B}}=1$, $\gamma^2\Delta=0.1$, $\sigma=3.5$, $\varepsilon_*=1$, $M=15$, $R_0=13$, $\lambda=1$.}
    \label{arctanplot}
\end{figure}

\begin{figure}
    \centering
    \captionsetup{justification=raggedright}
    \includegraphics[width=0.9\textwidth]{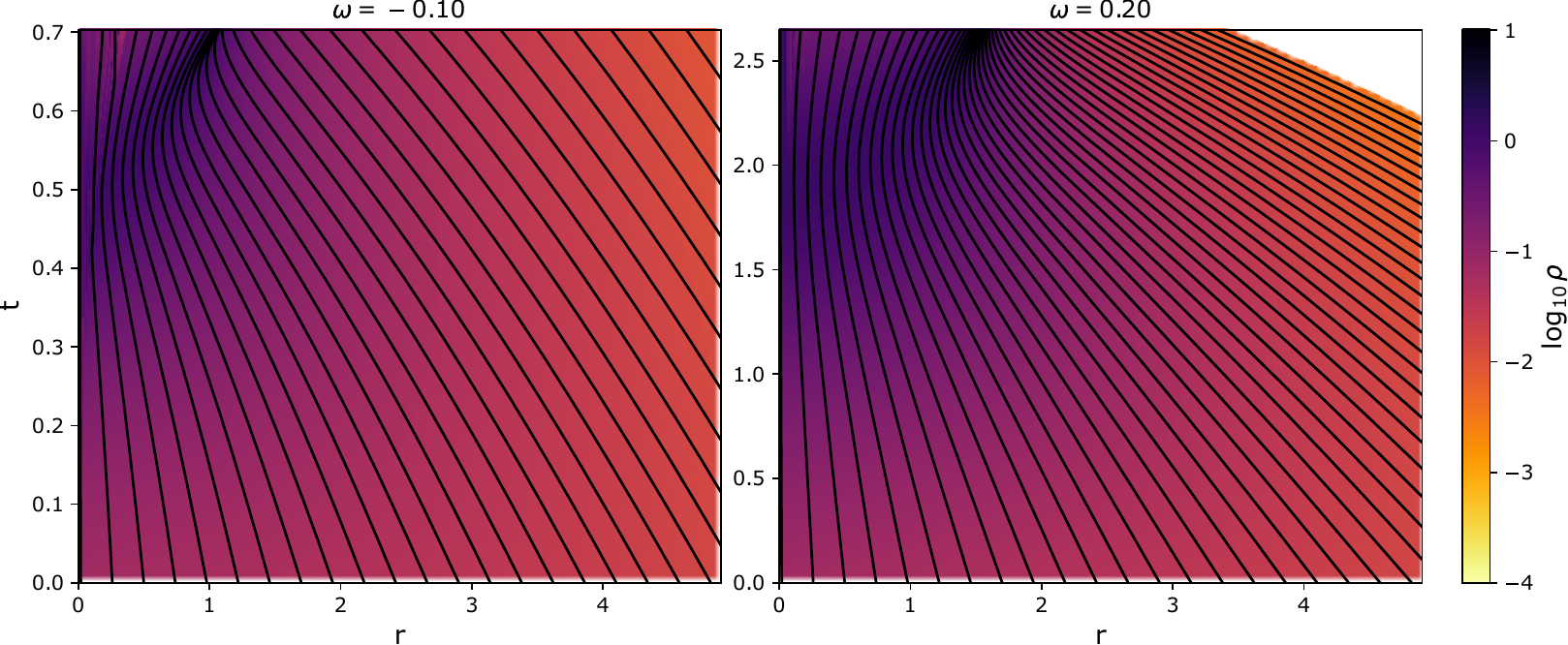}
    \caption{Shells dynamics and shell-crossing singularity formation for $\rho_5$ in \cref{rho5} with $\omega=-0.1$ (left) and $\omega=0.2$ (right). Natural units and constants are set to $c=G=k_{\mathrm{B}}=1$, $\gamma^2\Delta=0.1$, $\sigma=3.5$, $\varepsilon_*=1$, $M=30$, $R_0=11$, $\lambda=1.8$.}
    \label{FermiDiracplot}
\end{figure}

We also investigated the two profiles  
\begin{align}
&\rho_4 (t_0,R)= C_4 \left[\pi/2-\arctan\left(\frac{{R}-{R}_0}{\lambda}\right)\right] ,
\label{rho4}
\end{align}
\begin{align}
& C_4 =\frac{M}{{\int_{0}^{R_B}4 \pi \Tilde{R}^2 \left[\pi/2-\arctan\left(\frac{\tilde{R}-R_0}{\lambda}\right)\right] \dd \Tilde{R}}}\,,
\end{align}
and   
\begin{align}
\rho_5 (t_0,R)=\frac{C_5}{1+\exp\Big(\frac{R-R_0}{\lambda}\Big)}\, ,
\label{rho5}
\end{align}
\begin{align}
& C_5=\frac{M}{\int_{0}^{R_B} 4 \pi \Tilde{R}^2 \Big[1+\exp\big(\frac{\tilde{R}-R_0}{\lambda}\big)\Big]^{-1}\dd \Tilde{R}}\,.
\end{align}
The results of the numerical simulations are shown in Figs.~\ref{arctanplot} and~\ref{FermiDiracplot}.
These plots show the exact same qualitative physics we found for the density profile $\rho_3$ in Fig.~\ref{tanhplot}. Shell-crossing singularities develop just after the quantum bounce for both positive and negative pressure, and they do so in the ``tail'' of the density distribution. The inner shells bounce regularly. 

In addition, the previous and other profiles were studied in extreme pressure regimes, where $\omega$ is close or equal to $-1/3$ and $1$. We found a generic occurrence of shell crossings (not SCS) in regions close to the boundary of the star and before the quantum bounce. Even though these events are not physical singularities, the numerical integration breaks down in these events as well. This means that we were not able to probe the physics of the quantum region for these extreme pressures.

\begin{figure}
    \centering
    \captionsetup{justification=raggedright}
    \includegraphics[width=0.9\textwidth]{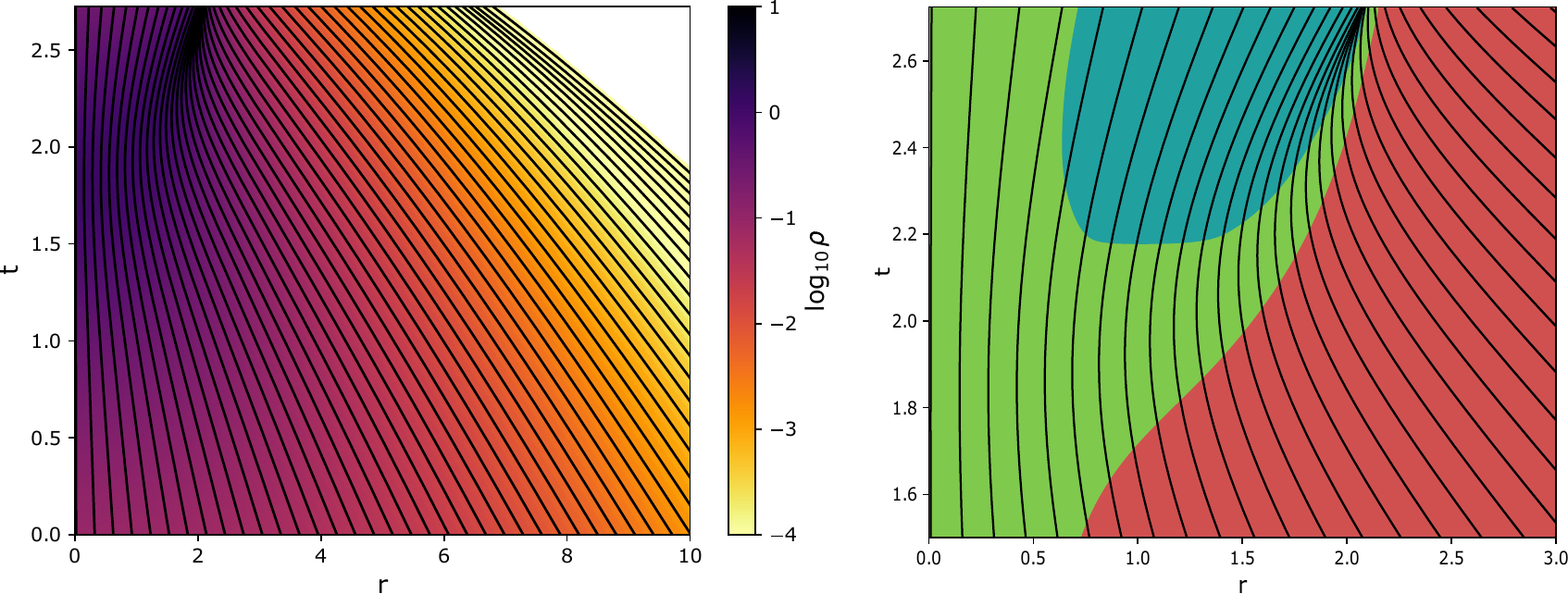}
    \caption{On the left: shells dynamics and shell-crossing singularity formation for $\rho_6$ in \cref{rho6} with $\omega=0.1$. On the right: section of the plot on the left with the trapped region highlighted in red, the non-trapped region in green, and the anti-trapped region in blue; the boundaries of these regions give the dynamical trajectories of the zeros of $\theta_+\theta_-$. Natural units and constants are set to $c=G=k_{\mathrm{B}}=1$, $\gamma^2\Delta=0.1$, $\sigma=3.5$, $\varepsilon_*=1$, $M=60$, $\lambda=3.5$.}
    \label{disegno}
\end{figure}

Another interesting feature of stellar collapse spacetimes to study before the formation of a shell-crossing singularity is the creation and disappearance of trapped and anti-trapped regions. We do so for the last density profile we consider:
\begin{align}
\rho_6 (t_0,R)=\frac{C_6}{\sqrt{2 \pi \lambda^2}} \exp\left(- \frac{R^2 }{ 2 \lambda^2}\right) , 
\label{rho6}
\end{align}
\begin{align}
& C_6=\frac{M}{\int_{0}^{R_B} 4 \pi \frac{\Tilde{R}^2}{ \sqrt{2 \pi \lambda^2} } \exp\left(- \frac{\tilde{R}^2 }{ 2 \lambda^2}\right)\dd \Tilde{R}}\,. 
\end{align}
The qualitative physics of the trapped regions of the previous profiles is similar to the one shown for $\rho_6$. Given an outgoing/ingoing null radial vector
\begin{equation}
    n_{\pm}=n^0 \Big( \partial_t \pm \frac{N \sqrt{1+\varepsilon}}{r^{\prime}} \,\partial_R \Big),
\end{equation}
the null expansions $\theta_{\pm}$ are given by $\theta_{\pm}=\nabla_{\mu} n_{\pm}^{\mu}$. Trapped, anti-trapped, and non-trapped regions are then delimited by the zeros of $\theta_{\pm}$, or, more conveniently, by the zeros of
\begin{equation}
    \theta_+ \theta_- \propto 1-\frac{\dot{r}^2}{N^2 (1+\varepsilon)}\,.
\end{equation}
The numerical simulation for the profile $\rho_6$ is shown in Fig.~\ref{disegno}. The qualitative physics of the shell dynamics and SCS formation (left panel) is once again equivalent to the one discussed for the previous density profiles. We can however gather some more information from the plot of the trapped regions on the right panel. In the latter, the trapped region is shown in red, the non-trapped region in green, and the anti-trapped region in blue. As we chose initial conditions such that a black hole already exists, at early times we find one zero of the null expansion, namely one trapping horizon, separating a trapped and a non-trapped region. As the collapse evolves, this trapping horizon moves outward. Then, after the bounce of the star's core, an anti-trapped region forms, and the shells inside of it are forced to move outwards. Soon after the anti-trapped region is formed, a shell-crossing singularity develops at the boundary of the anti-trapped and non-trapped regions, resulting in the breakdown of the equations of motion in the differential form.

\section{Vacuum solutions}
\label{sec:vacuum}

Vacuum solutions for the LQG effective equations in the dust-time gauge were studied in~\cite{Cafaro:2024vrw, Cipriani:2024nhx}. Importantly, Birkhoff’s theorem no longer holds at the semiclassical level. Classically, the spherically symmetric solutions to the vacuum Einstein field equations consist only of the static 1-parameter family of Schwarzschild spacetimes of mass $M$. At the semiclassical level, solutions to the effective equations with different $\varepsilon$ become non-diffeomorphic. This leads to a static 2-parameter family of Schwarzschild-like solutions of mass $M$ and parameter $\varepsilon(t,r)=\varepsilon=\mathrm{const}$. to the vacuum LQG effective equations in spherical symmetry. Furthermore, there exist even more solutions, where $\varepsilon(t,r)$ remains a function of both $t$ and $r$, that are presumably non-stationary. The relevance and physical significance of these last solutions are still unknown.

The situation is however even more complex than this. As we will show shortly using the effective equations of \cref{sec:effectiveeqs}, there also exist further vacuum spherically-symmetric semiclassical static solutions that cannot be put in the dust-time gauge, and so that could not be found from the previous analyses~\cite{Cafaro:2024vrw, Cipriani:2024nhx}. Consider in fact the line element in generalized Painlevé-Gullstrand coordinates (see \cref{PG1})
\begin{equation} 
\dd s^2=-N^2 \dd t^2 +\frac{1}{1+\varepsilon} (\dd r + N^r \dd t)^2 +r^2 \dd \Omega^2 .
\label{PGvacuum}
\end{equation}
The metric is defined by the three functions $N(t,r)$, $N^r (t,r)$, and $\varepsilon(t,r)$. The easiest way to look for static solutions of the vacuum ($\rho=\Pi=\Sigma=0$) effective equations of motion is to search for solutions where these functions are time-independent. Let us start from \cref{eps_dot} for $\varepsilon(t,r)$. We have $\partial_t \varepsilon =0$, and so $\varepsilon=\varepsilon(r)$, if either
\begin{equation} 
    \frac{\partial_r N}{N} = \frac{\partial_r \varepsilon}{2 (1+\varepsilon)} \, ,
    \label{N_eps}
\end{equation}
or $\sin(\frac{\sqrt{\Delta} b}{r})=0$, or $\cos(\frac{\sqrt{\Delta} b}{r})=0$. The last two options lead to trivial solutions~\cite{Cafaro:2024vrw, Cipriani:2024nhx}, so we focus on \cref{N_eps}. The solution to this equation reads $N(t,r)=f(t)\sqrt{1+\varepsilon(r)}$, where $f(t)$ is an integration function that can be reabsorbed in the choice of $t$. This means that $\partial_t \varepsilon =0$ leads to $\partial_t N =0$:
\begin{equation} 
N(r)=\sqrt{1+\varepsilon(r)}\,.
    \label{N_eps2}
\end{equation}
The last quantity we have to deal with is the shift $N^r (t,r)$. We could look for time-independent solutions of \cref{shiftvector,b_dot}. However, it is much more straightforward to use \cref{veryimportant} and, since we are now in vacuum, set $m$ to
\begin{equation}
M=4\pi \int_0 ^{r_B} \rho ~\Tilde{r}^2 \dd \Tilde{r}=\mathrm{const}.,
\label{Mvac}
\end{equation}
with $r_B$ the physical radius of the star at $t_0$. As the right-hand side of \cref{veryimportant} is now explicitly time-independent, so is $N^r = N^r (r)$. This gives an explicitly stationary solution to the vacuum semiclassical equations of motion, whose line element in generalized Painlevé-Gullstrand coordinates reads
\begin{align}
    \dd s^2 &=-f(r) \, \dd t^2 +\frac{\dd r^2}{1+\varepsilon(r)} +\frac{N^r (r)}{1+\varepsilon(r)} \, 2\, \dd r \dd t + r^2 \,d\Omega^2 , \label{vacuum_PG}\\
    f(r) & \coloneqq 1+\varepsilon - \frac{(N^r)^2}{1+\varepsilon} = 1- \frac{2 G M}{r} + \frac{\gamma^2 \Delta}{r^2} \left( \frac{2 G M}{r} + \varepsilon \right)^2 ,
    \label{fvacuum}
\end{align}
and the associated Killing vector field $\xi$ is given by $\xi=\partial_t$. It is straightforward to change coordinates to the LTB ones and to obtain the line element in \cref{LTB} with $N^2=1+\varepsilon(r(t,R))$. As these coordinates are not adapted to the Killing symmetry, the metric functions remain time-dependent in this gauge. Another interesting coordinate system for the vacuum solutions is the Schwarzschild one $(t_s,r)$, where the Schwarzschild time coordinate $t_s$ is defined by
\begin{equation}
    \dd t=\dd t_s+\frac{N^r}{(1+\varepsilon) \,f} \, \dd r_s\,.
\end{equation}
The line element in this coordinates takes the usual form
\begin{equation} \label{vacuum_SCH}
    \dd s^2=-f(r)\, \dd t_s^2 + f^{-1}(r) \,\dd r^2 + r^2 \, \dd \Omega^2 \, .
\end{equation}
From \cref{fvacuum} we see that in the classical limit $\Delta \rightarrow 0$, this family of static solutions gives exactly the 1-parameter family of Schwarzschild spacetimes of mass $M$. At the semiclassical level, besides the parameter $\Delta$ controlling the scale of the quantum corrections, there is a new arbitrary function $\varepsilon (r)$ entering the family of static solutions. This is a huge enlargement of this family. These solutions were not found in previous analyses of the vacuum and spherically-symmetric effective equations because they cannot in general be put in the dust-time gauge. From \cref{N_eps,N_eps2} we see that for $\varepsilon(r)=\mathrm{const}$., also $N(r)=\mathrm{const}$. and so the latter can be reabsorbed in the definition of the time coordinate to obtain the dust-time gauge, in agreement with the previous results~\cite{Cafaro:2024vrw, Cipriani:2024nhx}.

Interestingly, as these solutions should be matched with the boundary of the collapsing star, the function $\varepsilon$ can be interpreted as a ``quantum hair'' that leaks information regarding the interior of the black hole to the exterior vacuum region. A more in-depth analysis of these solutions, and their relevance for stellar collapse, is left for future work.


\section{Summary and conclusions}
\label{sec:conclusions}

In this work, we have studied stellar collapse models with pressure within the framework of effective loop quantum gravity. We have confirmed the resolution of the classical black hole singularity already seen in the dust case and its replacement with a quantum bounce of the star, taking place when its energy density becomes of Planckian order. We have furthermore shown that, independently from the initial density profile chosen, the magnitude of the pressure, and the type of fluid considered, shell-crossing singularities generally develop in the semiclassical collapse of a star with pressure. These findings generalize previous semiclassical results based on dust and they give a central role to the physics of shell-crossing singularities formation and evolution in the gravitational collapse of a star.

The effective equations of motion for stellar collapse with pressure were derived in the Lemaître-Tolman-Bondi gauge, as the latter is particularly suitable for numerical analyses. We found these equations for the general class of anisotropic fluids with no heat conduction and viscosity. Although anisotropies of this kind are believed to be in fact relevant for the physics of some stars~\cite{cadogan2024}, we did not try to model these effects here. We used the freedom in the class of fluids considered to study the case of a collapsing star with vanishing radial pressure. Despite being clearly unphysical, this toy model allowed us to study analytically several interesting physical features of a stellar collapse model with pressure. We were able to explore in particular the interplay of quantum corrections and (tangential) pressure in the formation of turning points in the shells' trajectories, giving a lower bound for the areal radius at which they occur, and the role of the initial conditions, especially the initial kinetic energy, in the evolution of the collapse. The effective equations for the collapse evolution were finally solved numerically, and shell-crossing singularities were found to be a general feature of the model.

We then moved to the more realistic case of a perfect fluid source. Given the complexity of the problem, this case was studied mainly numerically. As a first step into stellar collapse models with pressure, we considered a linear equation of state. However, the numerical code we used can be easily generalized to more sophisticated equations of state and we plan to explore these possibilities in future work. We investigated the collapse of many initial configurations, both for negative and positive pressures, within the limits imposed by the energy conditions, and found shell-crossing singularities in almost all scenarios. The only cases where we did not find SCSs were the ones where a shell-crossing ($r'=0$, but $m'=0$ as well and $\rho$ remains finite) developed before a shell-crossing singularity, and the numerical simulation could not be continued any further.

It is a common belief that shell-crossing singularities would be resolved by either the inclusion of pressure, quantum gravitational corrections, or a combination of the two. We have shown that the most simple model including both, namely considering only the quantum gravitational corrections included in the LQG effective models and a linear equation of state for the star, is still not enough to resolve these (weak) singularities. This result gives a central role to the physics of shell-crossing singularities in semiclassical stellar collapse. It in fact shows that it is not sufficient to naively include pressure and quantum corrections to resolve them. In order to do so, it is instead necessary to consider accurately their physics to obtain more realistic equations of state and the relevant quantum corrections.

On the other hand, even if we assume shell-crossing singularities to be eventually resolved, we expect them to turn into a regular phenomenon like acoustic waves. It is then reasonable to assume that the somewhat singular dynamics in the future of the shell-crossing singularities, which can be found by looking for weak solutions to the equations of motion as done in~\cite{Nolan_2003}, gives a qualitatively accurate description of the regular dynamics of these acoustic waves. A first step in this direction was taken in~\cite{Husain:2022gwp}, where weak solutions for semiclassical dust collapse were numerically investigated. This is however a complex topic, and a better understanding of these solutions, especially their causal structure, is still needed. In fact, the quantum gravitational corrections entering these semiclassical models neglect quantum phenomena taking place close to the horizon of the black hole. These include Hawking evaporation and any tentative scenario for what happens at the end of the evaporation, like the black-to-white hole transition~\cite{Haggard:2014rza,Christodoulou:2016vny,Bianchi:2018mml,Martin-Dussaud:2019wqc,DAmbrosio:2020mut,Soltani:2021zmv,Han:2023wxg,Han_2024}. These, together with the shockwaves resulting from the shell-crossing singularities, are all phenomena that, in the absence of a full quantum theory of the gravitational field, need to be studied somewhat separately and then put together coherently to obtain a comprehensive picture of the physical spacetime of a black hole. It is then paramount to have good control of the causal properties of all these phenomena in order to study the global causal structure of spacetime.

Finally, in the last section of this work, we briefly commented on the status of Birkhoff's theorem in the semiclassical setting. It was recently shown that, at the semiclassical level, solutions of the vacuum and spherically symmetric effective equations consist of a 2-parameter family $(M,\varepsilon)$ of static spacetimes, whose classical limit gives Schwarzschild spacetime, and also presumably non-stationary spacetimes~\cite{Cafaro:2024vrw, Cipriani:2024nhx}. We showed that the static solutions are even more general than we thought, as $\varepsilon$ in these solutions can be an arbitrary function of the areal radius $\varepsilon(r)$. As these further static solutions cannot be put in the dust-time gauge, they were overlooked by the previous analyses.

\acknowledgments
We thank Ed Wilson-Ewing for his valuable comments. The work of F.F. is supported in part by the Natural Sciences and Engineering Research Council of Canada. The work of F.S. and Lu.C. was funded by the National Science Centre, Poland, as part of the OPUS 24 grant number 2022/47/B/ST2/02735.

\appendix
\section{Conserved quantity in stellar collapse with pressure} \label{conservedquantity}

\Cref{epsilon,rhodotset} can be combined to obtain:
\begin{equation}
    \frac{\Dot{\varepsilon}}{2(1+\varepsilon)}-\frac{\Dot{\rho}}{\rho+\Pi}-\frac{\Dot{r}'}{r'}-2\frac{\Dot{r}}{r}\frac{\rho + \Sigma}{\rho + \Pi}=0  \, .
    \label{cq}
\end{equation}
Upon choosing linear EOSs, $\Pi=\omega_r \rho$ and $\Sigma=\omega_t \rho$, with $\omega_r$ and $\omega_t$ two different constants, \cref{cq} can be rewritten as
\begin{equation}
    \partial_t \ln\left[\frac{\sqrt{1+\varepsilon}}{r^{2 \frac{1+\omega_t}{1+\omega_r}} r' \rho^{\frac{1}{1+\omega_r}} } \right] =0   \, .
    \label{cq2}
\end{equation}
This implies that
\begin{equation}
    \frac{\sqrt{1+\varepsilon}}{r^{2 \frac{1+\omega_t}{1+\omega_r}} r' \rho^{\frac{1}{1+\omega_r}} }=\frac{4\pi }{ Q(R)}
    \label{conservedQ}
\end{equation}
is a conserved quantity. Even though its physical meaning is not obvious in the general case, in the dust case ($\omega_r=\omega_t=0$) \cref{conservedQ} reads
\begin{equation}
Q(R)=4 \pi \frac{ \rho \, r^2 r'}{\sqrt{1+\varepsilon}} \, .
\end{equation}
Noticing that the proper mass $m_{\mathrm P} (t,R)$ inside the shell $R$ can be written as
\begin{equation} \label{propermass}
m_{\mathrm P} (t,R)=4 \pi \int_0^R \frac{\rho \, r^2 r'}{\sqrt{1+\varepsilon}}\dd\tilde{R} = \int_0^R Q(\tilde{R}) \dd\tilde{R} \, ,
\end{equation}
it follows that $m_{\mathrm P} (R)$ is time independent. So, in the dust case, the conserved quantity associated with \cref{cq2} can be taken to be the proper mass $m_{\mathrm P} (R)$.

In the perfect fluid case ($\omega_r=\omega_t=\omega\neq 0$), it reads 
\begin{equation}
 Q(R)=\frac{4\pi\rho^{\frac{1}{1+\omega}} r^2 r'}{\sqrt{1+\varepsilon}}=\rho^{\frac{-\omega}{1+\omega}} \partial_R m_P(R,t)   \label{newconservation} \, .
\end{equation}
This conserved quantity is used in \cref{sec:perfect} to simplify the numerical integration of the equations of motion for the collapse of a perfect-fluid star.


\section{Avoidance of the classical bounce in the \texorpdfstring{$\Pi=0$}{Π=0} case}

As mentioned in \cref{TANG_PRES}, \cref{tang_pres} has a classical (first square bracket) and a quantum contribution (curly bracket). For $r$ such that one of the two brackets vanishes, then a bounce occurs. As we are concerned with the quantum properties of the model, we seek some bounds on the initial kinetic energy $K$ and $\omega$ such that no classical bounce can occur. As will be shown, this can be done for shells away from the center; however, when $R$ approaches $\sqrt{\Delta}$, a classical bounce seems to be an inevitable feature of the model.

Given the full expression of $\varepsilon$
\begin{equation} \label{eps_K_B}
    \varepsilon(t,R)=-1+\left[1-\frac{2 G m}{R}+\frac{R^2}{2 \gamma^2 \Delta}\left( 1-\sqrt{1- 8 K\frac{\gamma^2 \Delta}{R^{4\omega+2}}} \right)\right]\left(\frac{r}{R}\right)^{4 \omega} \, ,
\end{equation}
classical bounces are ruled out if $\left(\frac{2 G m}{r} + \varepsilon \right) >0$. Moreover, away from the center, it must be $r_{q} \le r \le R$ , with $r_q$ being the quantum bouncing radius. In a few steps, one can get
\begin{equation} \label{diseq_K}
    \sqrt{1- 8 K\frac{\gamma^2 \Delta}{R^{4\omega +2}}}  <  1+ \frac{2 \gamma^2 \Delta}{R^2} \left(1-\frac{2 G m}{R} \right)-\frac{2 \gamma^2 \Delta}{R^2} \left( \frac{R}{r}\right)^{4 \omega} \left(1-\frac{2 G m}{r} \right) \, .
\end{equation}
The presence of a negative term in the $RHS$ makes it non trivial to understand whether such an expression is always positive or not. Since $R/r \ge 1$ and $\left(1 -2Gm/R \right) > \left(1 -2Gm/r \right)$, if $\omega<0$ then the $RHS$ is always positive. For $\omega>0$, the condition $RHS>0$ can be easily rearranged to 
\begin{equation} \label{C3}
\left(\frac{r}{R} \right)^{4\omega}>\frac{1-\frac{2Gm}{r}}{1-\frac{2Gm}{R}+\frac{R^2}{2\gamma^2 \Delta}} \, .
\end{equation}
Here we will assume $R>2Gm(R)$ and $r>2Gm$. For $r<2Gm$ the inequality is trivially satisfied for each $\omega$. Since both the right and left-hand sides are increasing in $r$, applying the logarithm to both the members preserves the sign of the inequality 
\begin{equation}
\label{diseq_w}
\omega <\frac{1}{4}\frac{\ln\left(1-\frac{2Gm}{r}\right) -\ln\left(1-\frac{2Gm}{R}+\frac{R^2}{2\gamma^2 \Delta} \right)}{\ln\left(\frac{r}{R} \right)}  \, .
\end{equation}
Given that the denominator is always negative ($r<R$), so must be the numerator to have the inequality satisfied for some $\omega>0$. This last condition is equivalent to 
\begin{equation}
\frac{1-\frac{2Gm}{r}}{1-\frac{2Gm}{R}+\frac{R^2}{2\gamma^2 \Delta}}<1    \, ,
\end{equation}
which, by \cref{C3}, is always satisfied.

This short analysis showed that for 
\begin{equation} \label{w_max}
    \omega<\displaystyle\min_{r<R}\Bigg\{ \frac{1}{4}\frac{\ln\left(1-\frac{2Gm}{r}\right) -\ln\left(1-\frac{2Gm}{R}+\frac{R^2}{2\gamma^2 \Delta} \right)}{\ln\left(\frac{r}{R} \right)}    \Bigg\} \eqcolon \omega_{max} (R)
\end{equation} 
the $RHS$ of \cref{diseq_K} is positive. Consequently, both sides of \cref{diseq_K} can be squared to give:
\begin{equation} \label{diseq_K_2}
    \frac{R^{4\omega+2}}{8 \gamma^2 \Delta}- \frac{R^{4\omega+2}}{8 \gamma^2 \Delta} \left[ 1+ \frac{2 \gamma^2 \Delta}{R^2} \left(1-\frac{2 G m}{R} \right)-\frac{2 \gamma^2 \Delta}{R^2} \left( \frac{R}{r}\right)^{4 \omega} \left(1-\frac{2 G m}{r} \right) \right]^2< K(R) \leq \frac{R^{4\omega+2}}{8 \gamma^2 \Delta}\, .
\end{equation}
Notice that the $R$ dependence of $\omega_{\max}$ implies that at fixed $\omega$ some shells may violate \cref{diseq_w}. In fact, the expression for $\omega_{max}$ is exactly the one in \cref{bounce_c} with $\delta=1$. For this function, it was discussed that an arbitrary small $R$ implies an arbitrary small $\omega_c$. Even though the discussion depends on the density profile through $m(R)$, we expect this behavior to be a general feature of $\omega_{max}(R)$ too.

That being said, if we restrict ourselves to $R \gg \sqrt{\Delta}$, then \cref{diseq_w} holds, and the $LHS$ of \cref{diseq_K_2} can be Taylor-expanded, giving
\begin{equation}
 \frac{R^{4\omega}}{2}\left[\left(\frac{R}{r} \right)^{4\omega}\left(1-\frac{2Gm}{r} \right)-\left(1-\frac{2Gm}{R}\right) \right]+O\left(\frac{\Delta}{R^2}\right)<K(R) \leq \frac{R^{4\omega+2}}{8 \gamma^2 \Delta} \, .
\end{equation}
The left hand-side is free of $\Delta$, signaling its non-quantum nature. Additionally, it admits a global maximum at $r=2 G m (1 + 1/4 \omega)$, so the bounds for $R \gg \sqrt{\Delta}$ can be finally rewritten as
\begin{equation}
\label{lastinequality}
 \frac{R^{4\omega}}{2}\left[\left(\frac{2 \omega R}{Gm (4\omega+1)^2} \right)-\left(1-\frac{2Gm}{R} \right)    \right] <K(R) \leq  \frac{R^{4\omega+2}}{8 \gamma^2 \Delta} \, ,
 \end{equation}
neglecting $O\left(\Delta/R^2\right)$. All things considered, we conclude by saying that for $R \gg \sqrt{\Delta}$, there exists a fairly wide range of $K$ and $\omega$, respectively satisfying \cref{lastinequality} and \cref{w_max}, such that no classical bounce is produced.


\end{document}